\documentclass[10pt]{article}
\usepackage{graphviz}

\usepackage{amsmath}
\usepackage{amssymb}

\usepackage{graphicx}

\usepackage{cite}

\usepackage{color}

\usepackage{setspace} 
\usepackage{ifthen}

\usepackage[boxed,noend]{algorithm2e}

\def\pterror#1{\errmessage{Parsetree ERROR: #1}}

\newdimen\pthgap\def\pthorgap#1{\pthgap=#1}
\newdimen\ptvgap\def\ptvergap#1{\ptvgap=#1}

\newbox\ptnodestrutbox\def\ptnodestrut{\unhcopy\ptnodestrutbox}
\newbox\ptleafstrutbox\def\ptleafstrut{\unhcopy\ptleafstrutbox}

\def\ptnodefont#1#2#3{\def\ptnodefn{#1}
  \setbox\ptnodestrutbox=\hbox{\vrule height#2 width0pt depth#3}}
\def\ptleaffont#1#2#3{\def\ptleaffn{#1}
  \setbox\ptleafstrutbox=\hbox{\vrule height#2 width0pt depth#3}}

\pthorgap{12pt}                         
\ptvergap{12pt}                         
\ptnodefont{\normalsize\rm}{11pt}{3pt}  
\ptleaffont{\normalsize\it}{11pt}{3pt}  

\newcount\ptdepth           
\newcount\ptn               
\newbox\ptm \newdimen\ptmx  
\newbox\pta \newdimen\ptax  
\newbox\ptb \newdimen\ptbx  
\newbox\ptc \newdimen\ptcx  
\newbox\ptx \newdimen\ptxx  
\newif\ifpttri              

\def\ptnext{\advance\ptn by 1 \ifcase\ptn
  \or \setbox\ptm=\box\ptx \ptmx=\ptxx \or \setbox\pta=\box\ptx \ptax=\ptxx
  \or \setbox\ptb=\box\ptx \ptbx=\ptxx \or \setbox\ptc=\box\ptx \ptcx=\ptxx
  \else \pterror{More than 3 daughters in (sub)tree}\fi}

\def\ptbegtree{\ptdepth=0}
\def\ptendtree
  {\ifnum\ptdepth>0 \pterror{Mismatched bracketing: too few ')'s!}\fi}

\def\ptbeg{\ifnum\ptdepth=0 \leavevmode\fi\begingroup
  \advance\ptdepth1 \ptn=0\pttrifalse}
\def\ptend{\ifnum\ptdepth=0 \pterror{Mismatched bracketing: too many ')'s!}
  \else\ptcons\endgroup\ifnum\ptdepth=0 \box\ptx\else\ptnext\fi\fi}

\def\ptnodeaux#1{\setbox\ptx=\hbox{#1}\ptxx=0.5\wd\ptx\ptnext}
\def\ptnode#1{\ptnodeaux{\ptnodefn\ptnodestrut #1}}
\def\ptleaf#1{\ptnodeaux{\ptleaffn\ptleafstrut #1}}

\def\pthoradjust#1{\ifcase\ptn
  \or \pthadjbox{\ptm}{#1} \or \pthadjbox{\pta}{#1}
  \or \pthadjbox{\ptb}{#1} \or \pthadjbox{\ptc}{#1}
  \else \pterror{More than 3 daughters in (sub)tree}\fi}
\def\pthadjbox#1#2{\setbox#1=\hbox{\box#1\kern#2}}

\def\ptcons
 {\ifnum\ptn=0 \ptconsz 
  \else
    \ifpttri\ptconstri\else
      \ifcase\ptn \or\ptconsm\or\ptconsma\or\ptconsmab\or\ptconsmabc \fi \fi
    \ptax=\ptxx \advance\ptax-\ptmx \ptbx=0pt
    \ifdim\ptax<0pt \ptbx=-\ptax\ptax=0pt\ptxx=\ptmx \fi
    \setbox\ptx=\vtop{\hbox{\kern\ptax\box\ptm}\nointerlineskip
                      \hbox{\kern\ptbx\box\ptx}}\fi
  \global\ptxx=\ptxx\global\setbox\ptx=\box\ptx}

\def\ptavg#1#2#3{#1=#2\advance#1#3#1=0.5#1}     
\def\ptadv#1#2{\advance#1#2\advance#1\pthgap}   

\def\ptconsz{\ptxx=0pt \setbox\ptx=\vtop{}}     

\def\ptconsm{\ptxx=0pt 
  \setbox\ptx=\hbox{\ptedge{1}{0}{}{}}}         

\def\ptconsma                                   
 {\ptxx=\ptax\setbox\ptx=\vtop{
    \hbox{\ptedge{1}\ptax{}{}}\nointerlineskip
    \hbox{\box\pta}}}

\def\ptconsmab                                  
 {\ptadv\ptbx{\wd\pta}\ptavg\ptxx\ptax\ptbx
  \setbox\ptx=\vtop{
    \hbox{\ptedge{2}\ptax\ptbx{}}\nointerlineskip
    \hbox{\box\pta\kern\pthgap\box\ptb}}}

\def\ptconsmabc                                 
 {\ptadv\ptbx{\wd\pta}\ptadv\ptcx{\wd\pta}%
  \ptadv\ptcx{\wd\ptb}\ptavg\ptxx\ptax\ptcx
  \setbox\ptx=\vtop{
    \hbox{\ptedge{3}\ptax\ptbx\ptcx}\nointerlineskip
    \hbox{\box\pta\kern\pthgap\box\ptb\kern\pthgap\box\ptc}}}

\def\ptconstri                                  
 {\ifcase\ptn\or\setbox\pta\hbox{\kern2\pthgap}\or
  \or\setbox\pta=\hbox{\box\pta\kern\pthgap\box\ptb}
  \or\setbox\pta=\hbox{\box\pta\kern\pthgap\box\ptb\kern\pthgap\box\ptc}\fi
  \ptxx=0.5\wd\pta
  \setbox\ptx=\vtop{
    \hbox{\ptedge{0}{0}{\wd\pta}{}}\nointerlineskip
    \box\pta}}

\newcount\pted          
\newcount\ptedm         
\newcount\pteda         
\newcount\ptedb         
\newcount\ptedc         
\newcount\ptedl         
\newcount\ptedh         
\newcount\ptedhs        
\newcount\ptedvs        
\newcount\ptedtemp      

\def\ptedge#1#2#3#4{\pted=#1%
  \pteda=#2\ifcase\pted\ptedb=#3\or\or\ptedb=#3\or\ptedb=#3\ptedc=#4\fi
  \ptedm=\pteda\advance\ptedm\ifcase\pted\ptedb\or\pteda\or\ptedb\or\ptedc\fi
  \divide\ptedm by 2
  \ptedh=\ptvgap\ptedtemp=\ptedm\advance\ptedtemp-\pteda\divide\ptedtemp by 6
  \ifnum\ptedh<\ptedtemp\ptedh=\ptedtemp\fi
  \unitlength=1sp%
  \begin{picture}(0,\ptedh)
    \ifnum\pted=3 \ptedput\ptedc\fi
    \ifnum\pted=1 \else\ptedput\ptedb\fi
    \ptedput\pteda
    \ifnum\pted=0 \ptedbot\fi 
  \end{picture}}

\def\ptedput#1{\ptedl=#1\advance\ptedl-\ptedm
  \ifnum\ptedl>0 \ptedslope\else
    \ptedl=-\ptedl\ptedslope\ptedhs=-\ptedhs\fi
  \ifnum\ptedhs=0 \ptedl=\ptedh\fi
  \put(\ptedm,\ptedh){\line(\ptedhs,-\ptedvs){\ptedl}}}

\def\ptedbot
 {\ptedtemp=\ptedl\multiply\ptedtemp by \ptedvs\divide\ptedtemp by \ptedhs
  \ifnum\ptedtemp>0\ptedtemp=-\ptedtemp\fi \advance\ptedtemp-1
  \advance\ptedtemp\ptedh\multiply\ptedl by 2
  \put(\pteda,\ptedtemp){\line(1,0){\ptedl}}}

\def\ptedslope
 {\ifnum\ptedl>\ptedh\ptedhs=\ptedl\ptedvs=\ptedh
  \else              \ptedvs=\ptedl\ptedhs=\ptedh \fi
  \divide\ptedhs by 60 \divide\ptedvs by \ptedhs
  \ifnum \ptedvs <  5 \ptedvs=0 \ptedhs=1 \else
  \ifnum \ptedvs < 11 \ptedvs=1 \ptedhs=6 \else
  \ifnum \ptedvs < 13 \ptedvs=1 \ptedhs=5 \else
  \ifnum \ptedvs < 17 \ptedvs=1 \ptedhs=4 \else
  \ifnum \ptedvs < 22 \ptedvs=1 \ptedhs=3 \else
  \ifnum \ptedvs < 27 \ptedvs=2 \ptedhs=5 \else
  \ifnum \ptedvs < 33 \ptedvs=1 \ptedhs=2 \else
  \ifnum \ptedvs < 38 \ptedvs=3 \ptedhs=5 \else
  \ifnum \ptedvs < 42 \ptedvs=2 \ptedhs=3 \else
  \ifnum \ptedvs < 46 \ptedvs=3 \ptedhs=4 \else
  \ifnum \ptedvs < 49 \ptedvs=4 \ptedhs=5 \else
  \ifnum \ptedvs < 55 \ptedvs=5 \ptedhs=6 \else
                      \ptedvs=1 \ptedhs=1 
    \fi \fi \fi \fi \fi \fi \fi \fi \fi \fi \fi \fi
  \ifnum \ptedl < \ptedh
    \ptedtemp=\ptedhs \ptedhs=\ptedvs \ptedvs=\ptedtemp \fi}

\newenvironment{parsetree}{\ptactivechardefs\ptbegtree}{\ptendtree}

\def\ptcatcodes
 {\catcode`(=\active \catcode`)=\active
  \catcode`.=\active \catcode``=\active
  \catcode`~=\active}

{\ptcatcodes
\gdef\ptactivechardefs
 {\ptcatcodes
  \def({\ptbeg\ignorespaces}
  \def){\ptend\ignorespaces}
  \def.##1.{\ptnode{##1}\ignorespaces} 
  \def`##1'{\ptleaf{##1}\ignorespaces}
  \def~{\pttritrue\ignorespaces}}}

\usepackage{hyperref}

\makeatletter
\renewcommand{\@biblabel}[1]{\quad#1.}
\makeatother

\usepackage{xr}
\date{}

\newcommand\titlestring{Accurate reconstruction of insertion-deletion histories by statistical phylogenetics }
\newcommand\shorttitlestring{Insertion-deletion phylogenetics}

\usepackage{setspace}
\usepackage{array}

\usepackage{xr}
\newcommand{\appref}[1]{Appendix~\ref{app.#1}}
\newcommand{\applabel}[1]{\label{app.#1}}

\newcommand{\figref}[1]{Figure~\ref{Figures.#1}}
\newcommand{\figlabel}[1]{\label{Figures.#1}}

\newcommand{\eqnref}[1]{Equation~\ref{Equations.#1}}
\newcommand{\eqnlabel}[1]{\label{Equations.#1}}

\newcommand\argmax{\mbox{argmax}}

\newcommand\gappedalphabet[1]{({#1} \cup \{\epsilon\})}

\newcommand\gappedpair[2]{\gappedalphabet{#1} \times \gappedalphabet{#2}}

\newcommand\wtrans[4]{#1(#2 : [#3] : #4)}
\newcommand\transequiv{\equiv}

\newcommand\compose{}
\newcommand\identity{{\cal I}}

\newcommand\recognize{\nabla}

\newcommand\fork{\circ}

\newcommand\forkfun[2]{(#1) \fork (#2)}

\newcommand\alignments{alignments}

\newcommand\States{\Phi}
\newcommand\statesof[1]{\States_{#1}}

\newcommand\Transitions{\tau}

\newcommand\startstate{\phi_S}
\newcommand\laststate{\phi_E}

\newcommand\weight{{\cal W}}

\newcommand\numberofleaves{\kappa}

\newcounter{LeafIndex}
\newcommand\leafnode[1]{\numberofleaves \ifthenelse{\equal{#1}{1}}{}{\setcounter{LeafIndex}{#1} \addtocounter{LeafIndex}{-1} +\arabic{LeafIndex}}}

\newcommand\seqlen[1]{\mbox{length($#1$)}}

\newcommand\cardinality[1]{\left| #1 \right|}
\newcommand\order[1]{{\cal O}(#1)}

\newcommand\authorstring{
Oscar Westesson$^{1}$, 
Gerton Lunter$^{2}$, 
Benedict Paten$^{3}$, 
Ian Holmes$^{1,\ast}$
\\
\textbf{1} UC Berkeley and UCSF Graduate Program in Bioengineering, University of California, Berkeley, CA, USA; \\
\textbf{2} Wellcome Trust Center for Human Genetics, Oxford, Oxford, UK;\\
\textbf{3} Baskin School of Engineering, UC Santa Cruz, Santa Cruz, CA, USA
\\
$\ast$ E-mail: protpal@postbox.biowiki.org
}
\newcommand{\hl}[1]{#1}
\begin{document}

\begin{flushleft}
  {\Large
    \textbf{\titlestring}
  }
\\
\authorstring
\end{flushleft}
\newcommand\protpal{ProtPal}  
\newcommand\indelseqgen{indel-seq-gen}
\newcommand\xrate{XRate}
\newcommand\prank{PRANK}
\newcommand\muscle{MUSCLE}
\newcommand\mafft{MAFFT}
\newcommand\clustalw{CLUSTALW}
\newcommand\dawglambda{{\tt lambda.pl}}
\newcommand\dawg{DAWG}
\newcommand\indelign{Indelign}
\newcommand\fsa{FSA}
\newcommand\handalign{handalign}
\newcommand\baliphy{BAli-Phy}
\newcommand\probcons{ProbCons}
\newcommand\loyt{L\"oytynoja}
\newcommand{\e}[1]{\ensuremath{\times 10^{#1}}}
\newcommand\reconlink{\url{http://biowiki.org/~oscar/optic_reconstruction.tar}}
\newcommand\simulationlink{\url{http://biowiki.org/~oscar/simulation_reconstruction.tar}}

\newcommand{\rateParam}[1]{\lambda^{#1}}
\newcommand{\estimatedRate}[1]{\hat{\lambda}^{#1}_{\hat{H}}}
\newcommand{\rateRatio}[1]{\frac{\hat{\lambda}^{#1}_{\hat{H}}}{\lambda^#1}}

\newcommand{\lengthDist}[1]{{\bf p}^{#1}}
\newcommand{\subModel}{{\bf R}}

\section*{Abstract}
The Multiple Sequence Alignment (MSA) is a computational abstraction that represents a partial summary either of indel history,
or of structural similarity.
Taking the former view (indel history), it is possible to use formal automata theory
to generalize the phylogenetic likelihood framework for finite substitution models
(Dayhoff's probability matrices and Felsenstein's pruning algorithm)
to arbitrary-length sequences.
In this paper, we report results of a simulation-based benchmark
of several methods for reconstruction of indel history.
The methods tested include a relatively new algorithm for statistical marginalization of MSAs
that sums over a stochastically-sampled ensemble of the most probable evolutionary histories.
For mammalian evolutionary parameters on several different trees,
the single most likely history sampled by our algorithm
appears less biased than histories reconstructed by other MSA methods.
The algorithm can also be used for alignment-free inference, where the MSA is explicitly summed out of the analysis.
As an illustration of our method, we discuss reconstruction of the evolutionary histories of human protein-coding genes.

%

\tableofcontents

\section{Introduction}
The Multiple Sequence Alignment (MSA), indispensable to computational sequence analysis,
represents a hypothetical claim about the homology beteen sequences.
MSAs have many different uses, but the underlying hypothesis can often be classified as
a claim either of {\em structural} homology (the 3D structures align in a particular way)
or of {\em evolutionary} homology (the sequences are related by a particular history on a given phylogenetic tree).
These types of hypothesis are similar, but with subtle (and important) distinctions:
at the residue level, a claim of evolutionary homology (direct shared descent)
is far stronger than a claim of structural homology (same approximate fold).
Furthermore, both types of MSA---evolutionary and structural---typically only represent {\em summaries} of the respective homologies:
some fine detail is often omitted.
For example, an evolutionary MSA may---or may not---include the ancestral sequences at internal nodes of the underlying tree.

Structural and evolutionary MSAs are often conflated, but they have quite different applications.
For example, a common use for a structural MSA is {\em template-based structure prediction},
where a query sequence is aligned to a target of known structure;
the success of this prediction reflects the number of query-template residues correctly aligned \cite{QuEtAl2009}.  
By way of contrast, a common application for an evolutionary MSA is to identify regions or sites under selection,
the success of which depends on accurate reconstruction of the evolutionary history \cite{MosesEtAl2004, PollardEtAl2006}.

Evaluation of alignment methods is typically done with implicit regard for the structural interpretation.
Many benchmarks have used metrics based on the Sum of Pairs Score (SPS) \cite{ThompsonEtal99}.
In the situation that a query-template pairwise alignment is randomly picked out of the MSA,
the SPS effectively estimates the proportion of homologous residues that are correctly identified.
Several alignment methods attempt to maximize the posterior expectation of SPS or similar metrics.
This appears to improve accuracy, particularly when measured with reference to structural homology.
However, it does not automatically confer {\em evolutionary} accuracy --- a correct reconstruction of the evolutionary history of the sequences.  

Several studies suggest that multiple alignment for evolutionary purposes is still a highly uncertain procedure \cite{WongEtAl2008}, and that errors therein may significantly bias analyses of evolutionary effects \cite{LoytynojaGoldman2008,MarkovaRaina2011,NelesenEtAl2008,LiuEtAl2009,MarguliesEtAl2007,BradleyUzilovEtAl2009}.
%
A useful component of these studies is simulation of genetic sequence evolution \cite{LoytynojaGoldman2008},
which appears to better indicate evolutionary accuracy than benchmarks derived from protein structure alignments.
Simulations can be made quite realistic given the abundance of comparative sequence data \cite{StropeEtAl2009}.

The current state-of-the-art in phylogenetic alignment software
is a choice between (on the one hand)
programs that lack explicit models of the underlying evolutionary process,
and so are not framed as statistical inference problems \cite{LoytynojaGoldman2008},
and (on the other hand) Bayesian Markov chain Monte Carlo (MCMC) methods, which are statistically exact but prohibitively slow \cite{HolmesBruno2001,SuchardRedelings2006}.

A telling observation is that while substitution rate is routinely measured from MSAs and used as an indicator of natural selection,
there is relatively little analogous use of indel rate.
As we report here, it seems highly likely that even if indel rate is a useful evolutionary signal (which is eminently plausible),
the present alignment methods distort measurements of this rate so far as to make it meaningless (see especially \figref{estimates_biases}).

In this paper, we frame phylogenetic sequence alignment as an approximate maximum likelihood (ML) inference.
Our inference algorithm assumes that the tree is known,
requiring a separate tree estimation protocol.
While this is a strong assumption,
it is in principle shared among all progressive    aligners (e.g. PRANK\cite{LoytynojaGoldman2005}, Muscle \cite{Edgar2004b}, ClustalW\cite{LarkinEtAl2007}, MAFFT\cite{KatohEtAl2005}).
The alignment-marginalized likelihoods reported by our algorithm allow for statistical
tests between alternative trees, and the functionality to estimate an initial alignment and guide
tree from unaligned sequences exists elsewhere in the DART package. 
Our framing uses automata-theoretic methods from computational linguistics to unify several previously-disjoint areas of bioinformatics:
Felsenstein's pruning algorithm for the phylogenetic likelihood function \cite{Felsenstein81},
progressive multiple sequence alignment \cite{HigginsEtal92},
and alignment ensemble representation using partial order graphs \cite{LeeGrassoSharlow2002}.
Our algorithm may be viewed as a stochastic generalization of pruning to infinite state spaces:
it retains the linear time and memory complexity of pruning (${\cal O}(NL)$ for $N$ sequences of length $L$),
while moderating the biasing effect of the MSA.
\hl{The algorithmic details of our method are outlined briefly in the Methods, 
and in more complete, mathematically precise terms (with a tutorial introduction) in a separately submitted work.}

Our software implementation of this algorithm is called \protpal.
We measured the accuracy of \protpal\ relative to leading non-MCMC alignment/reconstruction protocols
by simulating indels and substitutions on a known phylogeny,
withholding the true history and attempting to reconstruct it from the sequences at the tips of the tree.
The results show that all previous approaches to the reconstruction of ancestral sequences introduce significant biases,
including systematic underestimation of insertions and overestimation of deletions.
This contradicts previous claims that advances in the statistical foundations of alignment tools, supported by improvements in protein-structure benchmarks,
necessarily improve the accuracy of evolutionary parameter estimates like the indel rate \cite{Cartwright2005,LoytynojaGoldman2008,BradleyEtAl2009}.

\protpal\ introduces less bias than any other methods we tested, including \prank, the state-of-the-art phylogenetic progressive aligner \cite{LoytynojaGoldman2008}.  
Based on our tests, \protpal\ appears to be the best choice for small to moderately-sized analyses, such as a reconstruction of the history of proteins at the inter-species level in human evolutionary history.
Using \protpal\ to estimate indel rates for $\sim 7,500$ human protein-coding gene families, we find that per-gene indel rates are approximately gamma-distributed,
with 95\% of genes experiencing a mean rate of less than 0.1 indel events per synonymous substitution event.  We find that lengths of inserted and deleted sequences are comparably distributed, having medians 5 and 7, respectively.
The human lineage appears to have experienced unusually many insertions since the human-mouse split.
By mapping genes to Gene Ontology (GO) terms, we find that the 200 fastest-indel genes  are enriched for regulatory and metabolic functions.  
Possible applications \hl{and extensions} of our algorithm include phylogenetic placement, homology detection, and  reconstruction of structured RNA. 

\section{Results}

\subsection{Computational reconstruction of simulated histories}

We undertook to determine the ability of leading bioinformatics programs, including \protpal, to characterize mutation event histories.
We simulated indel histories on a tree, then attempted to reconstruct the MAP history, $\hat{H}$, using only knowledge of the sequences $S$ and the phylogeny $T$ (but not the sequence alignment).
\hl{The history $\hat{H}$ is the aligned set of observed extant and predicted 
ancestral sequences, 
such that insertion, deletion, and substitution events can be pinpointed to specific tree
branches (though not to specific time points on those branches).}

We then characterized the reconstruction quality both directly, by comparison of $\hat{H}$ to the true $H$,
and indirectly, by using $\hat{H}$ to estimate $\theta$, the evolutionary parameters:
\begin{equation}
\eqnlabel{mapHistoryParams}                      
\hat{\theta}_{\hat{H}} = \argmax_{\theta'} P(\theta'|\hat{H},S,T) = \argmax_{\theta'} P(\hat{H},S|T,\theta')
\end{equation}
where the latter step assumes a flat prior, $P(\theta') = \mbox{const.}$
We then compared the history-conditioned parameter estimate $\hat{\theta}_{\hat{H}}$ to the true $\theta$.  

This statistic is not without its problems.
For one thing, we use an initial guess of $\theta$ to estimate $\hat{H}$.
Furthermore, for an unbiased estimate, we should sum over all histories, rather than conditioning on the MAP reconstructed history.
This summing over histories would, however, require multiple expensive calculations of $P(S|T,\theta)$, where conditioning on $\hat{H}$ requires only one such calculation.  
Furthermore, parameter estimation conditioned on a MAP-reconstructed history is the {\em de facto} method employed by large-scale genomics studies focusing on indels \cite{KamnevaEtAl2010, ZhangEtAl2011,ZhuWangEtAl2009, GomezValeroEtAl2008}.

\subsubsection{Simulation model parameters}

The model parameters are $\theta = (\rateParam{i}, \rateParam{d}, \lengthDist{i}, \lengthDist{d}, \subModel)$:
the insertion and deletion rates ($\rateParam{i}, \rateParam{d}$),
indel length distributions ($\lengthDist{i}, \lengthDist{d}$)
and substitution rate matrix ($\subModel$).
Here we focus on the rates ($\rateParam{i}, \rateParam{d}$).


As described in \appref{datagen}
we generated data using an external simulation tool, \indelseqgen, varying insertion ($\lambda^i$), deletion ($\lambda^d$) and substitution rates ($r$)
over a range representative of per-gene rates in {\em Amniota} evolution (\figref{OPTIC_dist}). 
We varied indel rates (with $\lambda^i = \lambda^d$) between 0.005 and 0.08 expected indels per unit time, estimating that this range accounts for 95\% of human gene families.  We left the substitution model $(\subModel)$ and indel length distributions $(\lengthDist{i}, \lengthDist{d})$ fixed, employing \indelseqgen's empirically-estimated values.  

We performed simulations on mammalian, amniote and fruitfly phylogenies, using the taxa in those clades for which genomic sequence is actually available.
We found generally consistent results,
with common trends that were most pronounced on the largest of the three trees that we used
(the twelve sequenced {\em Drosophila} species \cite{ClarkEtAl2007}).
In discussing the trends, we will refer specifically to the results on this largest of the trees.

\subsubsection{Indel rate estimates}
\paragraph{Overall most accurate} We first set out to determine which program, when used to analyze a set of unaligned sequences, returns the indel rate estimate closest to the true rate.

We report the ratio of inferred rate to true rate for insertions $\rateRatio{i}$ and deletions $\rateRatio{d}$ in 
\figref{estimates_biases}, with each $\estimatedRate{*} \in \{ \estimatedRate{i}, \estimatedRate{d}\} $ defined as $\hat{\theta}_{\hat{H}}$ in \eqnref{mapHistoryParams}.  
No parameter estimate derived from a computationally reconstructed history approaches the level of accuracy achieved using the true history (labeled ``True simulated history'' in \figref{estimates_biases}). 

The results do not always concord with previous benchmarks that have measured accuracy using 3D structural alignments:
for example, the \fsa\ program, one of the most accurate aligners on structural benchmarks \cite{BradleyEtAl2009}, performs poorly here.
\hl{This discordance  may be due to the fundamental differences 
between evolutionary and structural homology, and the metrics used to assess each.
For instance, consider a region with many nearby and overlapping 
insertions and deletions.  
The spatial and temporal location of these insertion and deletion events
(in particular, the pinpointing of events to branches on the tree)
defines what the  ``perfect'' evolutionary reconstruction is.  
In contrast, even given perfect knowledge of the insertion/deletion history, a ``perfect''
structural alignment depends only on the proteins at the tips of the tree,
and this alignment could differ from the true  evolutionary reconstruction.

Fundamentally, the difference between FSA and } \protpal\ \hl{is the underlying metric that is being optimized
by each program:
FSA attempts to maximize a metric (AMA=Alignment Metric Accuracy)
which is essentially ``structural'' (in the sense that it predicts how many residues would
be correctly aligned in a pairwise alignment of two leaf-node sequences, as might be used
in structure prediction by target-template alignment),
while }\protpal\ \hl{attempts to maximize a ``phylogenetic'' metric
(the probability of a given evolutionary history).
The metric we have used in our benchmark
(which counts correct reconstruction of the number of indel events on branches of the tree)
is also ``phylogenetic''.
By contrast,} \appref{suppSim} \figref{ama} \hl{shows the programs' ranking using the AMA metric.  
FSA perfoms well, with accuracy exceeding that of} \protpal\ \hl{in the highest
 indel rate category. 
This suggests that the differences between our evolutionary benchmark and previous benchmarks
are not due to the data, but rather the types of metrics that are used to measure alignment accuracy;
similarly, the differences between the leading programs are primarily due to what types of benchmark they
are explicitly trying to perform well at.}


All programs other than \protpal\ display insertion-{\em versus}-deletion biases that are, to a varying degree, asymmetric.
Typically, the asymmetry is that insertions are underrepresented and deletions overrepresented. 
\protpal's bias, which is generally less than the other programs, is also the most symmetric: reconstructed insertions and deletions are roughly equally reliable, with both slightly underestimated.



Over the distribution of human gene rates used by this benchmark,
our phylogenetic likelihood approach, \protpal, provides the most accurate reconstructions of both insertion and deletion counts.
  \prank, which also uses a tree (but no likelihood), avoids  insertion-deletion biases to a certain extent, although insertion rates are slightly underestimated relative to deletion rates.
Since \protpal's MAP history estimation appears similar to the optimization algorithm of \prank, we suspect that \protpal's marginally better performance 
 is due primarily to its main difference in implementation: \protpal\ tracks  an {\em ensemble} of possible reconstructions during progressive tree traversal (Section \ref{methods}), whereas  \prank\ uses a single ``current best guess.''

\paragraph{Effect of indel rate variation} 
To investigate the effect of indel 
rate variation on estimation accuracy,
we separate each program's
error distributions by indel rate (\figref{byRate}).  We find that all programs' accuracy is strongly affected by the indel rate used in simulation.


As the true indel rate increases,
most programs' estimates  drift towards $\rateRatio{*} \to 0$.
This is consistent with the so-called ``gap attraction'' effect, where indels that are nearby in sequence
can be misinterpreted as substitution events \cite{Lunter2007b}.
Depending on the phylogenetic orientation of the events,
estimated rates can be elevated or lowered, with different biases for insertion and deletion rates (\figref{gap_attraction}).

Gap attraction and other biases operate simultaneously, and are sometimes opposed.
\muscle\ over-estimates the deletion rate under most conditions,
but (consistent with a trend where programs have lower $\rateRatio{*}$ at higher indel rates)
gets the deletion rate roughly correct in the highest-indel-rate category of our benchmark.
However, the alignments produced by \muscle\ at high indel rates are no more ``accurate'' by pairwise metrics (\appref{suppSim} \figref{ama}).
We conjecture that multiple, contradictory types of gap attraction are at work, e.g. Figures~\ref{Figures.gap_attraction}B and~\ref{Figures.gap_attraction}C.

After \protpal, the two most accurate reconstruction methods are \prank\ and \probcons\ (the latter combined with a parsimonious indel reconstruction).
\probcons\ produces more reliable insertion estimates than \prank\ in a broad range of benchmark categories, is tied with \prank\ for deletion estimates, and appears robust to indel rate variation.
\prank\ performs slightly better than \probcons\ in the slowest indel rate category we considered.
\protpal\ produces the most reliable estimates overall,
outperforming \probcons\ in all but the fastest indel rate category, and \prank\ in all but the slowest.
\paragraph{Sensitivity to substitution rate} 
As compared to variation of simulated indel rate, variation of simulated substitution rate appears to have little effect on the accuracy of indel reconstruction (\appref{suppSim} \figref{indelBarPlot_bySub}).
One notable exception is \fsa, which appears to be affected by the substitution rate more than the other programs.
For example, when the simulated indel and substitution rates are both low, \fsa\ is comparable to the most accurate of the other programs (\protpal);
but when the substitution rate is increased, \fsa's error is greater than the least accurate program (\clustalw). 
\hl{Errors in estimating  the substitution rate are comparable among 
the programs tested, and are similarly correlated with the simulation indel rate} (\appref{suppSim} \figref{subRateMatrix}).

\subsection{Reconstructed indel histories of human genes}

We present here a comprehensive set of reconstructions accounting for the evolutionary history of individual codons in human genes.  
We used genes in the {\bf O}rthologous and {\bf P}aralogous {\bf T}ranscripts in {\bf C}lades (OPTIC) database's {\em Amniota} set, comprised of the 5 mammals H. {\em sapiens}, M. {\em musculus}, C. {\em familiaris}, M. {\em domestica}, O. {\em anatinus}  and  G. {\em gallus} as an outgroup \cite{HegerPonting2008}.
Considering only those families with one unique ortholog per species (approximately 7,500 families), we combined tree branch statistics across genes, using the species tree in \appref{suppOptic} \figref{OPTIC.branches}.  
Our reconstructions are available at \reconlink, and we provide here  various graphical summaries of {\em Amniota} evolutionary history.
Several negative results stand in contrast to earlier-reported trends.

\paragraph{Indel rates} 
Insertion and deletion rates are approximately gamma-distributed (\figref{OPTIC_dist}).
Roughly 95\% of genes have indel rates $<0.1$ indels per synonymous substitution.  

\paragraph{Phylogenetic origins} 
In our simulations, \protpal\ pinpoints residues' ``branch of origin'' more reliably than other tools, with a 93\% accuracy rate (\appref{suppSim} \figref{ancestry}).
Many codons appeared to have been inserted following the human-mouse split (\appref{suppOptic} \figref{OPTIC.branch_indel_rates})

\paragraph{Branch-specific indel rates}
Using our reconstructions  to estimate the rates of indel mutations along specific tree branches, we find evidence of an elevated insertion rate in the human (black) branch, as well as on the the {\em Amniota - Australophenids} (pink) branch (\appref{suppOptic} \figref{OPTIC.branch_indel_rates}).  

\paragraph{Amino acid distributions}  Distributions over amino acids differ significantly between inserted, deleted and non-indel sequences (\appref{suppOptic}  \figref{OPTIC.dist}).  
In general, small residues are over-represented in insertions, in agreement with previous studies \cite{ChauxEtAl2007}. 
\paragraph{Indel lengths} 
We find, contrary to a previous study in {\em Nematode}  \cite{WangEtAl2009}, that length distributions in the Amniotes are nearly identical between insertions and deletions (\appref{suppOptic} \figref{OPTIC.indel_distribution}).  The previously-reported result  may be attributable to the deletion-biased nature of the methods used, particularly \clustalw\ and \muscle\ \cite{WangEtAl2009}.
\paragraph{Indel position} The position of indels within genes is highly biased towards the ends of genes, presumably in large part reflecting annotation error (\appref{suppOptic} \figref{OPTIC.indel_positions}).  The bias is strongest for deletions at the N-terminus of the gene, but both insertions and deletions are enriched in both C- and N- termini.

\paragraph{Evolutionary context of indel SNPs}
We find no general correlation between the indel rate for a gene and the number of indel polymorphisms recorded for that gene in dbSNP \cite{SacconeEtAl2011} (\appref{suppOptic} \figref{OPTIC.indelVsSNP}).

\paragraph{Gene ontology indel rates} 
No Gene Ontology (GO) categories stand out as having significantly lowered or heightened indel rates in any of the three ontologies,
contrasting with the reported results of a 2007 study using a smaller number of genes \cite{ChauxEtAl2007}.    
An enrichment analysis conducted with GOstat \cite{BeissbarthSpeed2004}
showed that the 200 fastest evolving genes in our data are significantly enriched for regulatory
and metabolic functions.

\section{Discussion}
We developed and analyzed a simulation benchmark that compares programs based on their reconstructions of evolutionary history,
using instantaneous mutation rates representative of 
Amniote evolution.
We tested several different tree topologies;
results were similar on all trees, but most pronounced on the tree with the longest branch lengths.
We find that most programs distort indel rate measurements,
despite claims to the contrary.
Moreover, the systematic bias varies significantly when the rates of substitutions and indels are varied within a biologically reasonable range.
Many of the programs we rated have been ranked in the past, but using benchmarks that use protein structural alignments as a gold standard,
rather than evolutionary simulations.
Furthermore, these previous benchmarks have not directly assessed the reconstruction of evolutionary history (or summary statistics such as the indel rate),
but have used other alignment accuracy metrics such as the {\em Sum of Pairs Score}.
Alignment programs that perform weakly on our benchmark have apparently performed well on these previous benchmarks.
We hypothesize that these benchmarks, compared to ours, are less directly predictive of a program's accuracy at historical reconstruction,
although they may better reflect the program's suitability to assist in tasks
relating more closely to folded structure,
like prediction of a protein's 3D structure from a homologous template.
We have introduced a new notation that describes a general, hidden Markov model-structured likelihood function for indel histories on a tree, as well as the structure of the corresponding inference algorithm.
We have implemented the new method in a freely-available program, \protpal, that allows, for the first time, phylogenetic reconstruction with accuracy over a broad range of indel rates.
  \protpal\ is written in C++ as a part of the DART package: \url{www.biowiki.org/ProtPal}. 
The evolutionary reconstructions \protpal\  produces are, according to our simulated tests, the most accurate of any available tool,
for a range of parameters typical of human genes.

We applied \protpal\ to the reconstruction of human gene indel history, using families of human gene orthologs from the OPTIC database.
We find some patterns that agree with previous studies, such as the non-uniform distributions over amino acids seen in \cite{ChauxEtAl2007}. 
Other results stand in contrast - a previous study found significantly different  length distributions for insertions and deletions \cite{WangEtAl2009}, whereas in our data they appear very similar. 
Another prediction of our reconstruction  is an elevated rate of insertions on the human branch since the human-mouse split.  This contrasts with a previous analysis \cite{MouseGenome}, though the data therein was whole genomes, rather than individual protein-coding genes. 
In contrast to \cite{ChauxEtAl2007}, we find no obvious predictive power of the Gene Ontology (GO) for indel rates;
that is, the indel rate does not appear strongly correlated with the presence or absence of any particular GO term-gene association.
However, enrichment analysis for GO terms using GOstat \cite{BeissbarthSpeed2004}
showed that the 200 fastest-evolving genes are significantly enriched for regulatory and metabolic function.  This apparent discrepancy might be explained by a group of regulatory and metabolic genes which have very high indel rates, but whose small number prevent them from skewing the average within their GO categories. 

Many applications which use a fixed-alignment phylogenetic likelihood could potentially benefit from \protpal's reconstruction profiles.  For example, phylogenetic placement algorithms estimate taxonomic distributions by evaluating the relative likelihoods of placing sequence reads on tree branches \cite{MatsenEtAl2010}.  By using sequence profiles exported from \protpal, these reads could be placed with greater attention to indels and a more realistic accounting for alignment uncertainty.
Homology detection could be done in a similar way, thereby making use of the phylogenetic relationship of the sequences within the reference family.  It has been observed that the  detection of  positive selection is highly sensitive to the alignment used \cite{MarkovaRaina2011}.  \protpal\ could be modified to detect selection using entire profiles rather than single alignments, potentially eliminating the bias brought on by an inaccurate alignment.

In summary, multiple alignments are frequently constructed for use in downstream evolutionary analyses.
However, except for our method and slow-performing MCMC methods, there are no software tools for reconstructing molecular evolutionary history that explicitly maximize a phylogenetic likelihood for indels.
Our results strongly indicate that algorithms such as \protpal\ (which use such a phylogenetic model)
produce significantly more reliable estimates of evolutionary parameters, which we believe to be highly indicative of  evolutionary accuracy.  
These results falsify previous assertions that existing, non-phylogenetic tools are well-suited to this purpose.
Furthermore, we have demonstrated that it is possible to achieve such accuracy without sacrificing asymptotic guarantees on time/memory complexity, or resorting to expensive MCMC methods.
\protpal\ can reconstruct phylogenetic histories of entire databases on commodity hardware, enabling the large-scale study of evolutionary history in a consistent phylogenetic framework.

\section{Methods}
\label{methods}
\hl{The details concerning generation and analysis of simulated data are contained in} \appref{datagen}. 
\hl{A mathematically complete description of the alignment algorithm has been submitted
as a separate work, and an early version has been made available online here:}
 \url{http://arxiv.org/abs/1103.4347}. 

\subsection{Felsenstein's algorithm for indel models}
Our algorithm may be viewed as a generalization of Felsenstein's pruning recursion \cite{Felsenstein81},
a widely-used algorithm in bioinformatics and molecular evolution.
A few common applications of this algorithm include estimation of substitution rates \cite{Yang94b};
reconstruction of phylogenetic trees \cite{RannalaYang96};
identification of conserved (slow-evolving) or recently-adapted (fast-evolving) elements in proteins and DNA \cite{SiepelHaussler04b};
detection of different substitution matrix ``signatures''
(e.g. purifying vs diversifying selection at synonymous codon positions \cite{YangEtAl2000},
hydrophobic vs hydrophilic amino acid signatures \cite{ThorneEtAl96},
CpG methylation in genomes \cite{SiepelHaussler04},
or basepair covariation in RNA structures \cite{KnudsenHein99});
annotation of structures in genomes \cite{SiepelHaussler04c,PedersenEtAl2006};
and placement of metagenomic reads on phylogenetic trees \cite{MatsenEtAl2010}.

Felsenstein's algorithm computes $P(S|T,\theta)$ for a substitution model by tabulating intermediate probability functions of the form $G_n(x) = P(S_n|x_n=x,\theta)$,
where $x_n$ represents the individual residue state of ancestral node $n$,
and $S_n$ represents all the sequence data that is causally descended from node $n$ in the tree (i.e. the observed residues at the set of leaf nodes whose most recent common ancestor is node $n$).

The pruning recursion visits all nodes in postorder.
Each $G_n$ function is computed in terms of the functions $G_l$ and $G_r$ of its immediate left and right children (assuming a binary tree):
\begin{eqnarray*}
G_n(x) & = & P(S_n|x_n = x,\theta) \\
& = & \left\{
\begin{array}{ll}
\left( \sum_{x_l} M^{(l)}_{x,\ x_l} G_l(x_l) \right) \left( \sum_{x_r} M^{(r)}_{x,\ x_r} G_r(x_r) \right) & \mbox{if $n$ is not a leaf}
 \\
\delta(x=S_n) & \mbox{if $n$ is a leaf}
\end{array}
\right.
\end{eqnarray*}
where $M^{(n)}_{ab} = P(x_n=b|x_m=a)$ is the probability that node $n$ has state $b$, given that its parent node $m$ has state $a$;
and $\delta(x=S_n)$ is a Kronecker delta function terminating the recursion at the leaf nodes of the tree.
These $G_n$ functions are often referred to as ``messages'' in the machine-learning literature \cite{KschischangEtAl98}.

Our new algorithm is algebraically equivalent to Felsenstein's algorithm,
if the concept of a ``substitution matrix'' over a particular alphabet is extended to the countably-infinite set of all sequences over that alphabet.
Our chosen class of ``infinite substitution matrix'' is one that has a finite representation:
namely, the {\em finite-state transducer}, a probabilistic automaton that transforms an input sequence to an output sequence,
and a familiar tool of statistical linguistics \cite{MohriPereiraRiley2000}.

By generalizing the idea of matrix multiplication ($AB$) to two transducers ($A$ and $B$),
and introducing a notation for feeding the same input sequence to two transducers in parallel ($A \fork B$),
we are able to write Felsenstein's algorithm in a new form (see Section~\ref{sec:PhylogeneticLikelihood}):
\[
G_n = \left\{
\begin{array}{ll}
\left( M^{(l)} G_l \right) \fork \left( M^{(r)} G_r \right) & \mbox{if $n$ is not a leaf} \\
\recognize(S_n) & \mbox{if $n$ is a leaf}
\end{array}
\right.
\]
where $\recognize(S_n)$ is the transducer equivalent of the Kronecker delta $\delta(x=S_n)$.
The function $G_n$ is now encapsulated by a transducer ``profile'' of node $n$.

This representation has complexity ${\cal O}(L^N)$ for $N$ sequences of length $L$, 
which we reduce to ${\cal O}(LN)$ by stochastic approximation of the $G_n$.
This approximation relies on the {\em alignment envelope} \cite{PatenEtAl2008},
a data structure introduced by prior work on efficient alignment methods.
The alignment envelope is a subset of all the possible histories
in which most of the probability mass is concentrated.
A related data structure is the {\em partial order graph} \cite{LeeGrassoSharlow2002}.
Both these data structures can be viewed as ensembles of possible histories, in contrast to a single ``best-guess'' reconstruction of the history.
\figref{sampledGraph} shows a state graph, with paths through it corresponding
to histories relating the two sequences GL and GIV.  The paths highlighted in blue form 
a partial order graph, corresponding to a subset of these histories generated by 
a stochastic traceback.  
At each progressive traversal step, we sample a high-probability subset of alignments of
two sibling profiles in order to maintain a bound on the state space size. 
Note that if we sample only the most likely path at every internal node, 
we essentially recover 
the progressive algorithm of PRANK, and if we sample and store all solutions, 
we recover the machine $G_n$ with state space of size ${\cal O}(L^N)$.

\subsection{Transducer definitions and lemmas}

The definitions and lemmas are presented in a condensed form here, and expanded upon in \cite{WestessonEtAlArxiv2011}.


A transducer is a tuple $(\Omega, \Psi, \States, \startstate, \laststate, \Transitions, \weight)$
where
$\Omega$ is an input alphabet,
$\Psi$ is an output alphabet,
$\States$ is a set of states,
$\startstate \in \States$ is the start state,
$\laststate \in \States$ is the end state,
$\Transitions \subseteq \States \times \gappedalphabet{\Omega} \times \gappedalphabet{\Psi} \times \States$ is the transition relation, and
$\weight:\Transitions \to [0,\infty)$ is the transition weight function.

Suppose that
 $T = (\Omega, \Psi, \States, \startstate, \laststate, \Transitions, \weight)$
and
 $U = (\Omega', \Psi', \States', \startstate', \laststate', \Transitions', \weight')$
are transducers.

Let $\weight(\pi)$ be the product of all transition weights along a state path $\pi$
and let $\wtrans{\weight}{x}{T}{y}$ be the sum of such weights for all paths whose
input labels, concatenated, yield the string $x \in \Omega^\ast$ and whose output labels yield $y \in \Psi^\ast$.

{\em Equivalence:}
If $T$ and $U$ have the same input and output alphabets ($\Omega=\Omega'$ and $\Psi=\Psi'$)
and the same sequence weights $\wtrans{\weight}{x}{T}{y}=\wtrans{\weight'}{x}{U}{y}\ \forall x,y$,
then we say the transducers are {\em equivalent}, $T \transequiv U$.
Less formally, we will write $T \cong U$ if $\wtrans{\weight}{x}{T}{y} \simeq \wtrans{\weight'}{x}{U}{y}$.

{\em Moore transducers:}
The {\em Moore normal form} for transducers, named for Moore machines \cite{Moore56},
associates input/output with three distinct types of state: {\em Match}, {\em Insert} and {\em Delete}.
Paths through Moore transducers can be associated with (gapped) pairwise alignments of input and output sequences.
For any transducer $T$, there exists an equivalent Moore-normal form transducer $U$
with $\cardinality{\States'} = \order{\cardinality{\Transitions}}$
and $\cardinality{\Transitions'} = \order{\cardinality{\Transitions}}$.

{\em Composition:}
If $T$'s output alphabet is the same as $U$'s input alphabet ($\Psi = \Omega'$),
there exists a transducer, $T\compose U = (\Omega, \Psi', \States'' \ldots \weight'')$, that unifies the output of $T$ with the input of $U$,
such that $\forall x \in \Omega^\ast, z \in (\Psi')^\ast$:
\begin{equation}\eqnlabel{composedWeight}
\wtrans{\weight''}{x}{T\compose U}{z} = \sum_{y\in\Psi^\ast} \wtrans{\weight}{x}{T}{y} \wtrans{\weight'}{y}{U}{z}
\end{equation}
If $T$ and $U$ are in Moore form,
then $\cardinality{\States''} \leq \cardinality{\States} \times \cardinality{\States'}$
and $\cardinality{\Transitions''} \leq \cardinality{\Transitions} \times \cardinality{\Transitions'}$.

{\em Intersection:}
If $T$ and $U$ have the same input alphabets ($\Omega = \Omega'$),
there exists a transducer, $T\fork U = (\Omega, \Psi'', \States'' \ldots \weight'')$, that unifies the input of $T$ with the input of $U$.
The output alphabet is $\Psi'' = \gappedpair{\Psi}{\Psi'}$,
i.e. a $T$-output symbol (or a gap) aligned with a $U$-output symbol (or a gap).

Let $\alignments(t,u) \subset (\Psi'')^\ast$ denote the set of all gapped pairwise alignments of sequences $t \in \Psi^\ast$ and $u \in (\Psi')^\ast$.
Transducer $T\fork U$ has the property that $\forall x \in \Omega^\ast, t \in \Psi^\ast, u \in (\Psi')^\ast$:
\begin{equation}\eqnlabel{forkWeights}
\sum_{v \in \alignments(t,u) } \wtrans{\weight''}{x}{T\fork U}{v} = \wtrans{\weight}{x}{T}{t} \wtrans{\weight'}{x}{U}{u}
\end{equation}
If $T$ and $U$ are in Moore form,
then $\cardinality{\States''} \leq \cardinality{\States} \times \cardinality{\States'}$
and $\cardinality{\Transitions''} \leq \cardinality{\Transitions} \times \cardinality{\Transitions'}$.
Paths through $T \fork U$ are associated with three-way alignments of the input sequence to the two output sequences.

{\em Identity:}
Let $\identity$ be a transducer that copies input to output unmodified, so $\identity \compose T \transequiv T \compose \identity \transequiv T$.

{\em Exact match:}
For any sequence $S \in \Omega^\ast$, there exists a Moore-form transducer $\recognize(S) = (\Omega, \emptyset, \States, \Transitions \ldots)$
with $\cardinality{\States} = \order{\seqlen{S}}$
and $\cardinality{\Transitions} = \order{\seqlen{S}}$,
 that rejects all input except $S$,
such that $\wtrans{\weight}{x}{\recognize(S)}{\epsilon} = 1$ if $x=S$, and $0$ if $x \neq S$.        
Note that $\recognize(S)$ outputs nothing (the empty string).

{\em Chapman-Kolmogorov transducers:}        
A transducer $T$ is {\em probabilistic} if $\wtrans{\weight}{x}{T}{y}$ represents a probability $P(y|x,T)$:   
that is, for any given input string, $x$, it defines a probability measure on output strings, $y$.

Suppose $T(t)$ is a function returning a probabilistic transducer of the form $(\Omega, \Omega, \States, \startstate, \laststate, \Transitions, \weight(t))$,   i.e. a transducer whose transition weight $\weight$ depends on an additional {\em time parameter}, $t$,
and which satisfies the transducer equivalence
$T(t) \compose T(t') \transequiv T(t+t')\ \forall t,t'$.

Then $T(t)$ gives the finite-time transition probabilities of a homogeneous continuous-time Markov process on the strings $\Omega^\ast$,
as the above transducer equivalence is a form of the Chapman-Kolmogorov equation.

If the state space of $T$ is finite, then this equation describes a renormalization of the composed state space $\States \times \States$ back down to the original state space $\States$.   
 So far, only one nontrivial time-dependent transducer is known that solves this equation exactly using a finite number of states: the TKF91 model \cite{ThorneEtal91}.


\subsection{The phylogenetic likelihood}
\label{sec:PhylogeneticLikelihood}
We rewrite the evidence, $P(S|T,\theta)$ for sequences $S$, tree $T$, and parameters $\theta$,
in the form $P(\{S_n: n \in {\cal L}\}|R,\{B_n\})$
where $\{S_n: n \in {\cal L}\}$ denotes the set of sequences observed at leaf nodes,
$\{B_n\}$ denotes the stochastic evolutionary processes occuring on the branches,
and $R$ denotes the probabilistic model for the sequence at the root node of the tree.

The root and branch transducers $(R,\{B_n\})$ represent an alternative view of the tree and parameters $(T,\theta)$.
The root transducer $R$ outputs from the equilibrium or other initial distribution of the process.
If $(p,c) \in T$ is a parent-child pair, then $B_c = B(T_{pc})$ is a time-dependent transducer parameterized by the branch length.
In practise, the branch transducers need not satisfy the Chapman-Kolmogorov equation
for the following constructs to be of use;
for example, the $\{B_n\}$ might be approximations to true Chapman-Kolmogorov transducers \cite{MiklosLunterHolmes2004}.

Let $R=(\emptyset,\Omega,\ldots)$ be a transducer outputting sequences sampled from the prior at the phylogenetic root.

Let $n$ be a tree node.
If $n$ is a leaf, define $F_n = \identity$.
Otherwise, let $(l,r)$ denote the left and right child nodes, and define
$F_n = \forkfun{B_l \compose F_l}{B_r \compose F_r}$
where $B_n=(\Omega,\Omega,\ldots)$ is a transducer modeling the evolution on the branch leading to $n$.

Diagramatically we can write $F_n$ as
\begin{parsetree}
 ( .. ( .$B_l$. ( .$F_l$. ~ ) ) ( .$B_r$. ( .$F_r$. ~ ) )  )
\end{parsetree}

The phylogenetic likelihood is then fully described by $F = R \compose F_{\mbox{root}}$.

Like $R$, transducer $F$ models a probability distribution over output sequences, but accepts only the empty string as an input sequence.
This empty input sequence is just a technical formality (transducers must have inputs); if we ignore it, we can think of $F$ and $R$ as hidden Markov models (HMMs), rather than transducers.
$R$ is an HMM that generates a single sequence,
$F$ a multi-sequence HMM that generates the whole set of leaf sequences.

Inference with HMMs often uses a dynamic programming matrix (e.g. the Forward matrix)
to track the ways that a given evidential sequence can be produced by a given grammar.

For our purposes it is useful to introduce the evidence in a different way,
by transforming the model to incorporate the evidence directly.
We augment the state space so that the model is no longer capable of generating any sequences {\em except} the observed $\{S_n\}$,
by composing $F_{\mbox{root}}$'s forked outputs with exact-match transducers that will only accept the observed sequences at the leaves of the tree.
This yields a model, $G$, whose state space is of size $\mathcal{O}(L^N)$ and, in fact, is directly analogous to the Forward matrix.

If $n$ is a leaf node, then let $G_n = \recognize(S_n)$ where $S_n$ is the sequence at $n$.
Otherwise, $G_n = \forkfun{B_l \compose G_l}{B_r \compose G_r}$.

Diagramatically we can write $G_n$ as
\begin{parsetree}
 ( .. ( .$B_l$. .$G_l$. ) ( .$B_r$. .$G_r$. )  )
\end{parsetree}

Let $G = R \compose G_{\mbox{root}}$.
The evidence is $P(\{S_n\}|R,\{B_n\}) = \wtrans{\weight}{\epsilon}{G}{\epsilon}$.

The net output of $G$ is always the empty string. The sequences $\{S_n\}$ are recognized as inputs by the $\recognize(S_n)$ transducers at the tips of the tree, but are not passed on as outputs themselves.

Likewise, the input of $G$ is the empty string, because $R$ accepts only the empty string on its input.

We can think of $G$ as a Markov model, rather than an HMM. It has no input or output; rather, the sequences are encoded into its structure.

Transducer $G$ has $\order{L^N}$ states, which is impractically many,
so \protpal\ uses a progressive hierarchy $H_n$ of approximations to the corresponding $G_n$, with state spaces that are bounded in size.

If $n$ is a leaf node, let $H_n = \recognize(S_n) = G_n$.
Otherwise, let $H_n = \forkfun{B_l \compose E_l}{B_r \compose E_r}$
where $\statesof{E_n} \subseteq \statesof{H_n}$
is a subset defined by sampling complete paths through the Markov model $M_n = R \compose H_n$
and adding the $H_n$-states used by those paths to $\statesof{E_n}$,
until the pre-specified bound on $\cardinality{\statesof{E_n}}$ is reached.
Then $G \cong M_{\mbox{root}}$.

The likelihood of a given history may be calculated by summing over paths through $G$ consistent with that history.
In the simplest cases (e.g. minimal Moore-form branch transducers), each indel history corresponds to exactly one path,
so the MAP indel history corresponds to the maximum-weight state path through $G$.

\subsection{Alignment envelopes}

Let $\recognize(S)$ be defined such that it has only one nonzero-weighted path
\[
X_0 \to W_0 \stackrel{S_1}{\to} M_1 \to W_1 \stackrel{S_2}{\to} M_2 \to \ldots \to W_{L-1} \stackrel{S_L}{\to} M_L \to W_L \to X_L
\]
so a $\recognize(S)$-state is either the start state ($X_0$), the end state ($X_L$), a wait state ($W_i$) or a match state ($M_i$).
All these states have the form $\phi_i$ where $i$ represents the number of symbols of $S$ that have to be read in order to reach that state,
i.e. a ``co-ordinate'' into $S$.
All $\recognize(S)$-states are labeled with such co-ordinates, as are the states of
any transducer that is a composition involving $\recognize(S)$,
such as $G_n$ or $H_n$.

For example, in a simple case involving a root node (1) with two children (2,3) whose sequences are constrained to be $S_2,S_3$,
the evidence transducer is $G = R \compose G_{\mbox{root}} = R \compose (G_2 \fork G_3) = R \compose (\forkfun{B_2 \compose \recognize(S_2)}{B_3 \compose \recognize(S_3)})$
=
\begin{parsetree}
 ( .$R$. ( .$B_2$. .$\recognize[S_2]$. ) ( .$B_3$. .$\recognize[S_3]$. )  )
\end{parsetree}

All states of $G$ have the form $g=(r,b_2,\phi_2 i_2,b_3,\phi_3 i_3)$ where $\phi_2, \phi_3 \in \{ X, W, M \}$,
so $\phi_2 i_2 \in \{ X_{i_2}, W_{i_2}, M_{i_2} \}$ and similarly for $\phi_3 i_3$.
Thus, each state in $G$ is associated with a co-ordinate pair $(i_2,i_3)$ into $(S_2,S_3)$, as well as a state-type pair $(\phi_2,\phi_3)$.

Let $n$ be a node in the tree,
let ${\cal L}_n$ be the set of indices of leaf nodes descended from $n$,
and let $G_n$ be the phylogenetic transducer for the subtree rooted at $n$,
defined in Section~\ref{sec:PhylogeneticLikelihood}.
Let $\States_n$ be the state space of $G_n$.

If $m \in {\cal L}_n$ is a leaf node descended from $n$,
then $G_n$ includes, as a component, the transducer $\recognize(S_m)$.
Any $G_n$-state, $g \in \States_n$, is a tuple, one element of which is a $\recognize(S_m)$-state, $\phi_i$, where $i$ is a co-ordinate (into sequence $S_m$) and $\phi$ is a state-type.
Define $i_m(g)$ to be the co-ordinate and $\phi_m(g)$ to be the corresponding state-type.

Let $A_n:\States_n \to 2^{{\cal L}_n}$ be the function returning the set of {\em absorbing leaf indices} for a state, such that the existence of a finite-weight transition $g' \to g$ implies that $i_m(g) = i_m(g') + 1$ for all $m \in A_n(g)$.

Let $(l,r)$ be two sibling nodes.
The {\em alignment envelope} is the set of sibling state-pairs from $G_l$ and $G_r$ that can be aligned.
The function $E\colon \States_l \times \States_r \to \{ 0,1 \}$ indicates membership of the envelope.
For example, this basic envelope allows only sibling co-ordinates separated by a distance $s$ or less

\begin{equation}\eqnlabel{basicEnvelope}
E_{\mbox{basic}}(f,g) = \max_{m \in A_l(f), n \in A_r(g)} |i_m(f)-i_n(g)| \leq s
\end{equation}

An alignment envelope can be based on a {\em guide alignment}. 
For leaf nodes $x,y$ and $1 \leq i \leq \seqlen{S_x}$, let $\mathcal{G}(x,i,y)$ be the number of residues of sequence $S_y$
in the section of the guide alignment from the first column, up to and including the column containing residue $i$ of sequence $S_x$.

This envelope excludes a pair of sibling states if they include a homology between residues which is more than $s$ from the homology of those characters contained in the guide alignment:
\begin{equation}\eqnlabel{guideEnvelope}
E_{\mbox{guide}}(f,g) = \max_{m \in A_l(f), n \in A_r(g)} \max(\ |\mathcal{G}(m,i_m(f),n)-i_n(g)| \ ,  |\mathcal{G}(n,i_n(g),m)-i_m(f)|\ ) \leq s
\end{equation}

Let $K(x,i,y,j)$ be the number of match columns (those columns of the guide alignment in which both $S_x$ and $S_y$ have a non-gap character) between the column containing residue $i$ of sequence $S_x$ and the column containing residue $j$ of sequence $S_y$.  
This envelope excludes a pair of sibling states if they include a homology between residues which is more than $s$ {\em matches} from the homology of those characters contained in the guide alignment:

\begin{align*}
E_{\mbox{guide}}(f,g) & = & \max_{m \in A_l(f), n \in A_r(g)} \max(\ |\mathcal{G}(m,i_m(f),n)-K(m,i_m(f),n,i_n(g))|, \ \\
& &|\mathcal{G}(n,i_n(g),m)-K(n,i_n(g),m,i_m(f))|\ ) \leq s
\end{align*}

\subsection{OPTIC data analysis}

\paragraph{Data} Amniote gene families were downloaded from \url{http://genserv.anat.ox.ac.uk/downloads/clades/}.  We restricted our analysis to the $\sim$ 7,500 families having simple 1:1 orthologies.  The same species tree topology (downloaded from \url{http://genserv.anat.ox.ac.uk/clades/amniota/displayPhylogeny} was used for all reconstructions, though branch lengths were estimated separately for each family as part of OPTIC.  When computing branch-specific indel rates, the branch lengths of the species tree were used.  

\paragraph{Reconstruction and rate estimation} Gene families were aligned and reconstructed using \protpal\ with a 3-rate class Markov chain over amino acids, insertion and deletion rates set to 0.01, and 250 traceback samples.  Averaged and per-branch indel rates were computed with \protpal\ using the {\tt -pi} and {\tt -pb}  options.  The indel rates were then normalized by the synonymous substitution rate for each corresponding nucleotide alignment (taken directly from OPTIC), computed with PAML \cite{Yang2007}. Residues' origins were determined by finding the tree node closest to the root containing a non-gap reconstructed character.  

\paragraph{External data}  Genes were mapped to Gene Ontology terms via the  mapping downloaded from \url{http://www.ebi.ac.uk/GOA/human\_release.html} during 10/2010.  Indel SNPs per gene were taken from a table downloaded from Supplemental Table 5 of  \cite{MillsEtAl2006}.

\section{Figures}
\begin{center}
\begin{figure}[h!]
\includegraphics[width=1\textwidth]{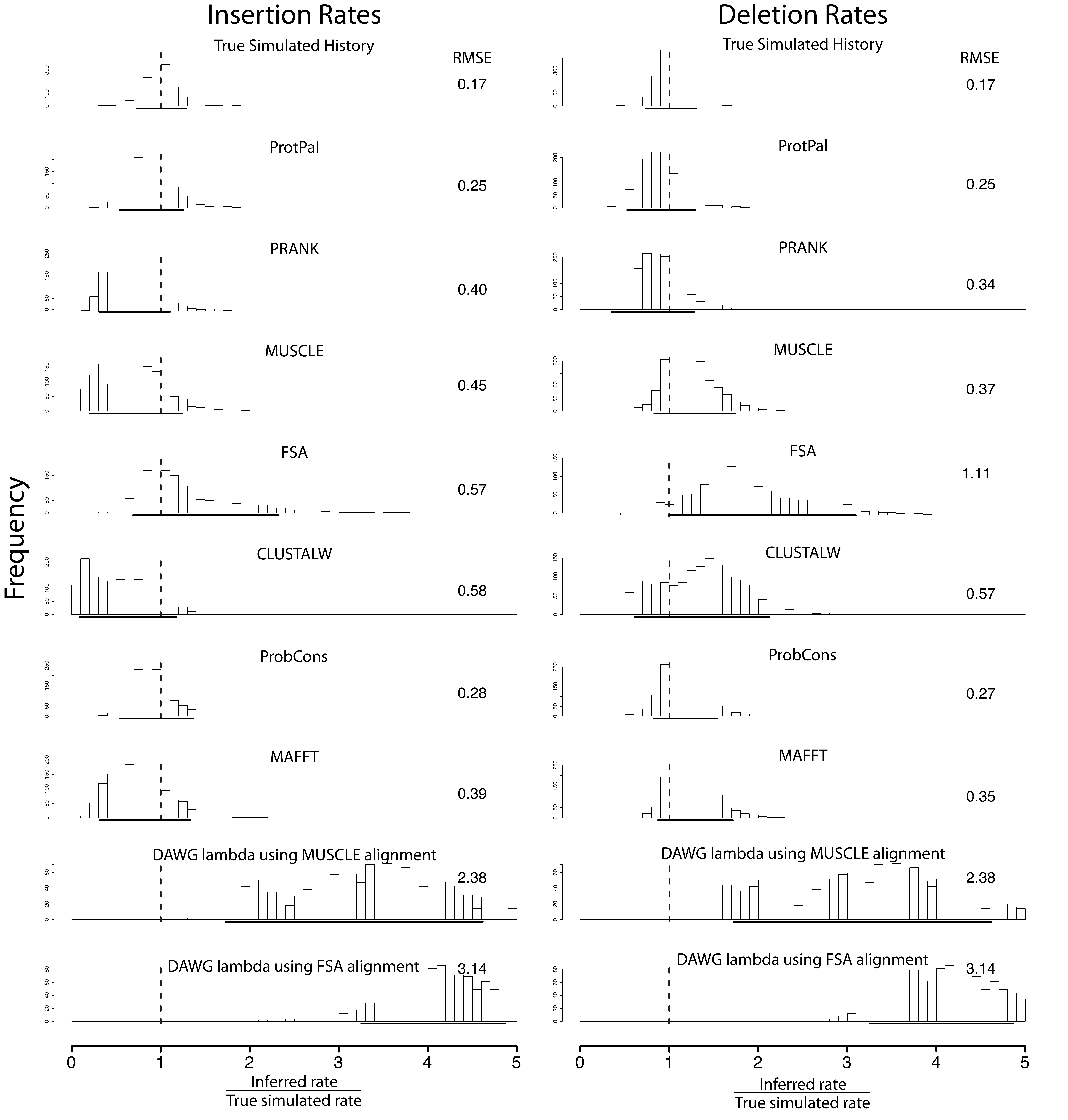}
\caption{ \protpal's estimates of insertion and deletion rates are the most accurate of any program tested, as measured by the RMSE of $\rateRatio{*}$ values aggregated over all substitution/indel rate categories. Quantiles containing 90\% of the data are shown as a bolded portion of the $x$-axis, and RMSE is shown to the right of each distribution, the latter computed as described in \appref{datagen} Equation 6.  No aligner approaches the accuracy of the rates estimated with the true alignment, though \protpal, \prank, and \probcons\ are the top three, with \protpal\ as the most accurate over all.  Many aligners, particularly \muscle, \clustalw, and \mafft, significantly underestimate insertion rates and overestimate deletion rates.  \protpal\ and \prank\ perform their own ancestral reconstruction and other alignment programs were augmented with a most-recent-common-ancestor (MRCA) parsimony as described in \cite{SinhaSiggia2005}.}
\figlabel{estimates_biases}
\end{figure}
\end{center}

\begin{center}
\begin{figure}[h!]
\includegraphics[width=1\textwidth]{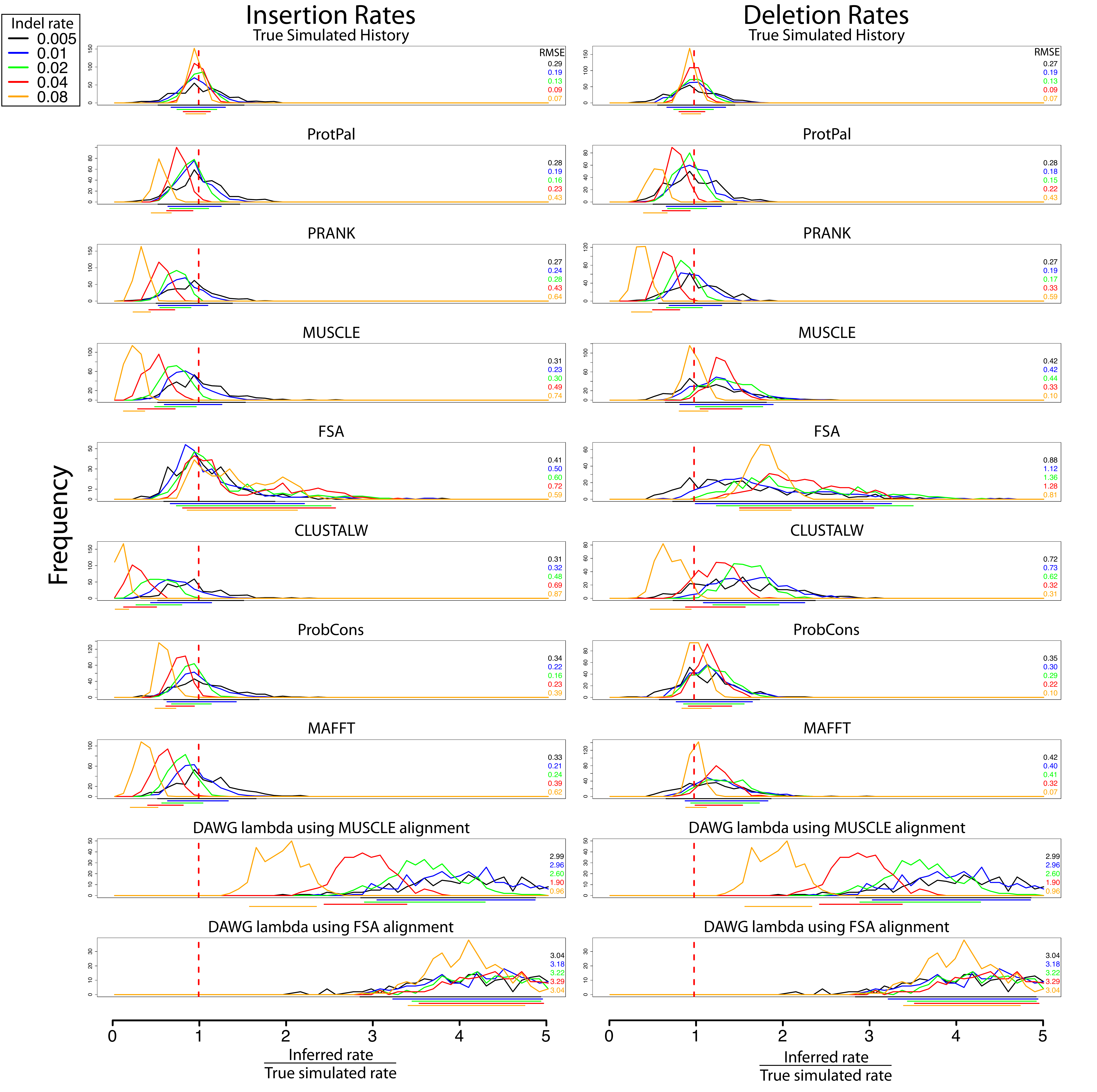}
\caption{
Rate estimation accuracy is highly dependent on the simulated indel rate.  For instance, \prank\ is more accurate at lower indel rates, \probcons\ is more accurate at higher rates.  \protpal\ is more accurate than \prank\ in all but one rate (0.005) and equal or more accurate than \probcons\ in all but one rate (0.08). The drift towards $\frac{inferred}{true}=0$ exhibited by most programs indicates that most programs infer proportionally fewer indels as rates are increased, likely due to various forms of gap attraction.  
Color-coded 90\% quantiles and RMSEs are shown underneath and to the right of each group of distributions, respectively.  RMSE is computed as described in  \appref{datagen} Equation 6}
\figlabel{byRate}
\end{figure}
\end{center}

\begin{center}
\begin{figure}[h!]
\includegraphics[width=1\textwidth]{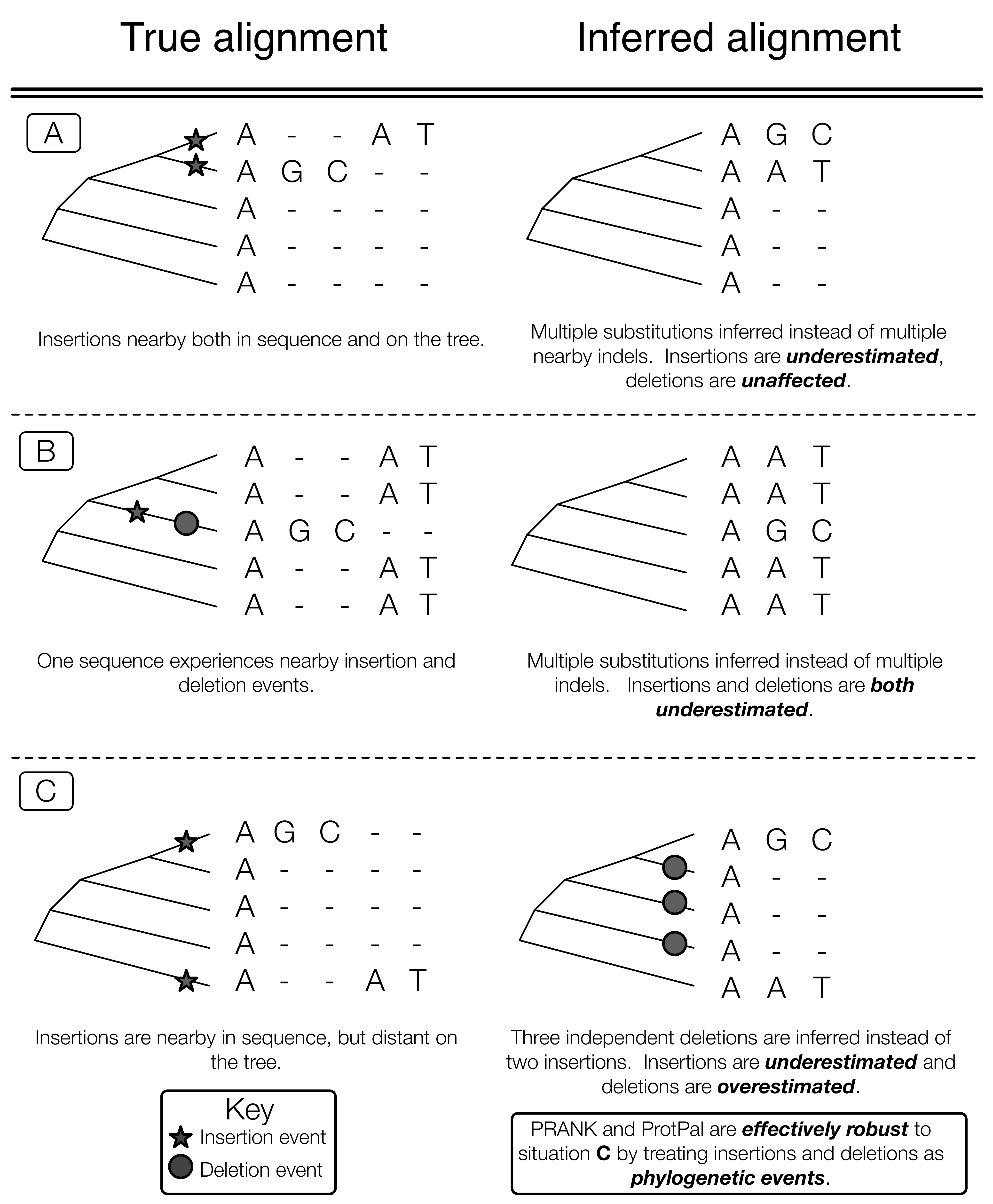}
\caption{
{\em Gap attraction}, the canceling of nearby complementary indels, can affect insertion and deletion rates in various ways depending on the phylogenetic relationship of the sequences involved.  All programs are, to some extent, sensitive to situations {\bf A} and  {\bf B} whereas phylogenetic aligners can avoid situation {\bf C}. 
An insertion at a leaf requires gaps at all other leaves - an understandably costly alignment move when gaps are added without regard to the phylogeny, resulting in {\bf multiple penalization} for each insertion.  Such a penalization would cause most non-phylogenetic aligners to prefer the ``Inferred alignment'' in case {\bf C} where there are fewer total gaps.  Aligners treating indels as phylogenetic events would penalize each of the implied multiple deletions and only penalize each insertion once, thus preferring the ``True alignment'' in case {\bf C}. 
}
\figlabel{gap_attraction}
\end{figure}
\end{center}

\begin{center}
\begin{figure}[h!]
\includegraphics[width=1\textwidth]{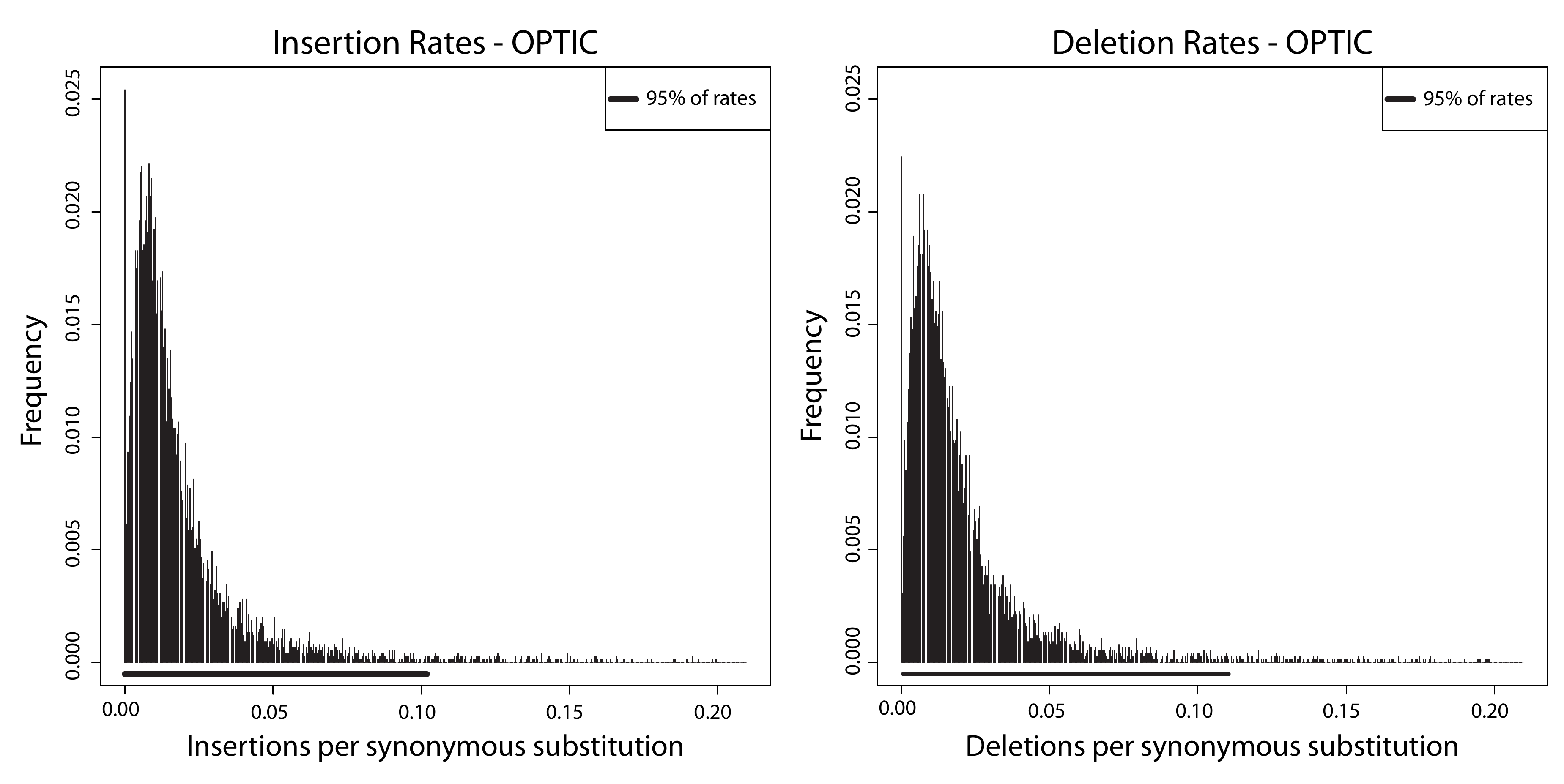}
\caption{
Insertion and deletion rates in {\em Amniota} show similar distributions, with 95\% of genes having rates less than approximately 0.1 indels per synonymous substitution.  
Insertion and deletion rates were estimated using reconstructions done with \protpal\ from a set of approximately 7,500 protein-coding genes from the OPTIC amniote database \cite{HegerPonting2008}.  Indel rates were normalized by the synonymous substitution rate of each gene as computed with PAML \cite{Yang2007} so that the plotted rate represents the number of expected indels per synonymous substitution.  
Since these rates are conditioned on the MAP reconstructed history,
there are many alignments whose inferred indel rates are zero (197, 174, and 54 for insertions, deletions, and both, respectively).
}
\figlabel{OPTIC_dist}
\end{figure}
\end{center}

\begin{center}
\begin{figure}[h!]
\includegraphics[width=1\textwidth]{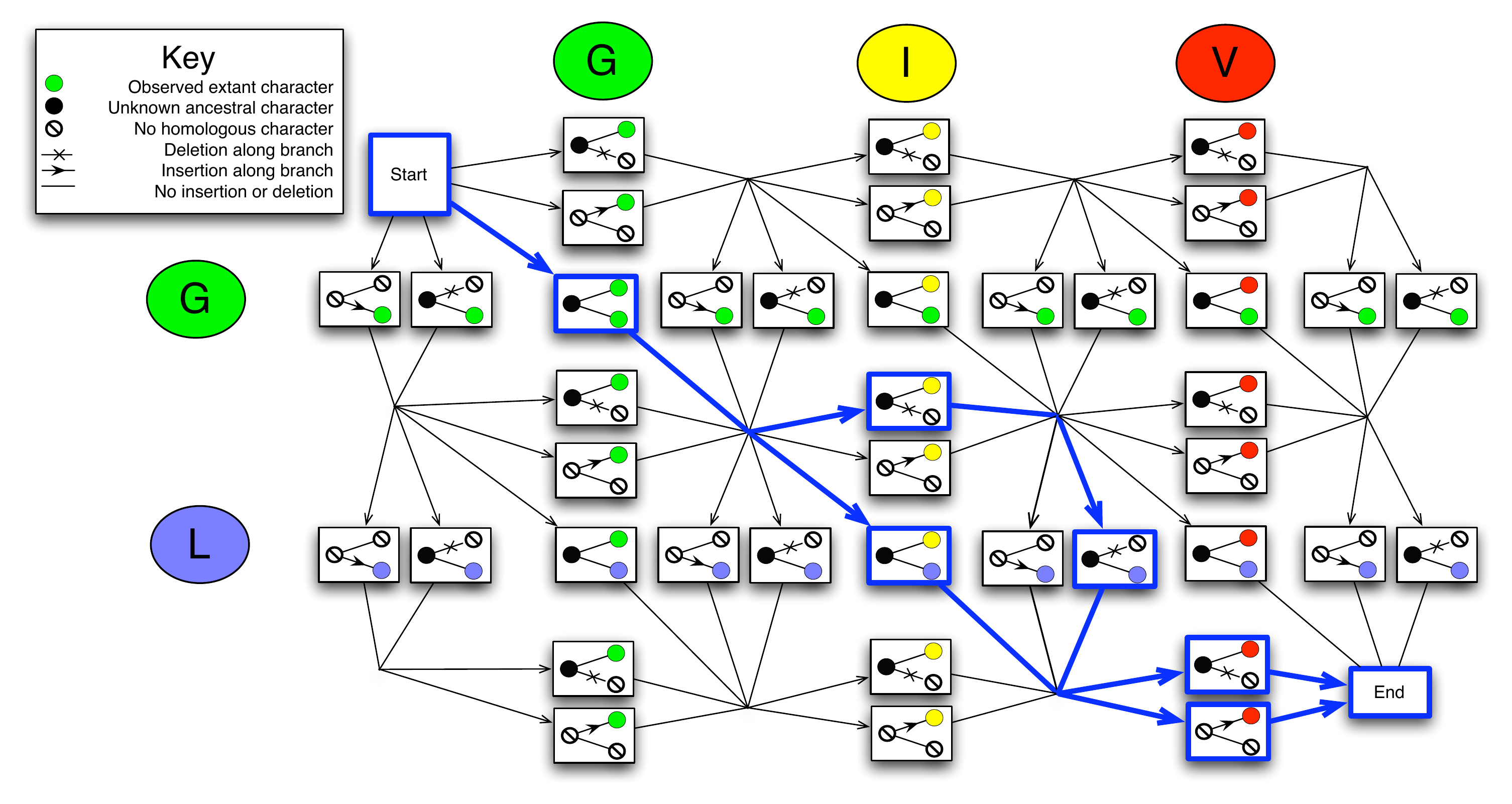}
\caption{Each  path through this state graph represents a possible evolutionary history
relating sequences GL and GIV.  By using stochastic traceback algorithms
 (sampling paths proportional
to their posterior probability, blue highlighted states and transitions), 
it is possible to select a high-probability subset of the full state graph.  
By constructing such a subset at each internal node, it is possible to maintain a bound
on the state space size during progressive  tree traversal 
 while still retaining an ensemble of possible histories. 
 }
\figlabel{sampledGraph}
\end{figure}
\end{center}

\pagebreak

 \section{References}
 \bibliography{../latex-inputs/alignment,../latex-inputs/ncrna,../latex-inputs/genomics,../latex-inputs/reconstruction}

\begin{thebibliography}{10}

\bibitem{QuEtAl2009}
X.~Qu, R.~Swanson, R.~Day, J.~Tsai, {\it Curr Protein Pept Sci.\/} {\bf 10},
  270 (2009).

\bibitem{MosesEtAl2004}
A.~M. Moses, D.~Y. Chiang, D.~A. Pollard, V.~N. Iyer, M.~B. Eisen, {\it Genome
  Biology\/} {\bf 5} (2004).

\bibitem{PollardEtAl2006}
K.~S. Pollard, {\it et~al.\/}, {\it Nature\/} {\bf 443}, 167 (2006).

\bibitem{ThompsonEtal99}
J.~D. Thompson, F.~Plewniak, O.~Poch, {\it Nucleic Acids Research\/} {\bf 27},
  2682 (1999).

\bibitem{WongEtAl2008}
K.~M. Wong, M.~A. Suchard, J.~P. Huelsenbeck, {\it Science\/} {\bf 319}, 473
  (2008).

\bibitem{LoytynojaGoldman2008}
A.~L\"oytynoja, N.~Goldman, {\it Science\/} {\bf 320}, 1632 (2008).

\bibitem{MarkovaRaina2011}
P.~Markova-Raina, D.~Petrov, {\it Genome Research\/} {\bf 21}, 863 (2011).

\bibitem{NelesenEtAl2008}
S.~Nelesen, K.~Liu, D.~Zhao, C.~R. Linder, T.~Warnow, {\it Pacific Symposium on
  Biocomputing\/} {\bf 2008}, 25 (2008).

\bibitem{LiuEtAl2009}
K.~Liu, S.~Nelesen, S.~Raghavan, C.~R. Linder, T.~Warnow, {\it IEEE/ACM Trans
  Comput Biol Bioinform\/} {\bf 6}, 7 (2009).

\bibitem{MarguliesEtAl2007}
E.~project consortium, {\it Genome Research\/} {\bf 17}, 760 (2007).

\bibitem{BradleyUzilovEtAl2009}
R.~K. Bradley, {\it et~al.\/}, {\it PLoS ONE\/} {\bf 4}, e6478 (2009).

\bibitem{StropeEtAl2009}
C.~Strope, K.~Abel, S.~Scott, E.~Moriyama, {\it Mol Biol Evol\/} {\bf 26}, 2581
  (2009).

\bibitem{HolmesBruno2001}
I.~Holmes, W.~J. Bruno, {\it Bioinformatics\/} {\bf 17}, 803 (2001).

\bibitem{SuchardRedelings2006}
M.~A. Suchard, B.~D. Redelings, {\it Bioinformatics\/} {\bf 22}, 2047 (2006).

\bibitem{LoytynojaGoldman2005}
A.~L\"oytynoja, N.~Goldman, {\it Proceedings of the National Academy of
  Sciences of the USA\/} {\bf 102}, 10557 (2005).

\bibitem{Edgar2004b}
R.~C. Edgar, {\it BMC Bioinformatics\/} {\bf 5}, 113 (2004).

\bibitem{LarkinEtAl2007}
M.~Larkin, {\it et~al.\/}, {\it Bioinformatics\/} {\bf 23}, 2947 (2007).

\bibitem{KatohEtAl2005}
K.~Katoh, K.~Kuma, H.~Toh, T.~Miyata, {\it Nucleic Acids Research\/} {\bf 33},
  511 (2005).

\bibitem{Felsenstein81}
J.~Felsenstein, {\it Journal of Molecular Evolution\/} {\bf 17}, 368 (1981).

\bibitem{HigginsEtal92}
D.~G. Higgins, A.~J. Bleasby, R.~Fuchs, {\it Computer Applications in the
  Biosciences\/} {\bf 8}, 189 (1992).

\bibitem{LeeGrassoSharlow2002}
C.~Lee, C.~Grasso, M.~Sharlow, {\it Bioinformatics\/} {\bf 18}, 452 (2002).

\bibitem{Cartwright2005}
R.~A. Cartwright, {\it Bioinformatics\/} {\bf 21 Suppl 3}, iii31 (2005).

\bibitem{BradleyEtAl2009}
R.~K. Bradley, {\it et~al.\/}, {\it PLoS Computational Biology\/} {\bf 5},
  e1000392 (2009).

\bibitem{KamnevaEtAl2010}
O.~Kamneva, A.~Liberles, N.~Ward, {\it Genome Biology and Evolution\/} {\bf 2},
  870 (2010).

\bibitem{ZhangEtAl2011}
Z.~Zhang, J.~Huang, Z.~Wang, L.~WAng, G.~Peiji, {\it Molecular Biology and
  Evolution\/} {\bf 28}, 291 (2011).

\bibitem{ZhuWangEtAl2009}
L.~Zhu, Q.~Wang, P.~Tang, H.~Araki, D.~Tian, {\it Molecular Biology and
  Evolution\/} {\bf 26}, 2353 (2009).

\bibitem{GomezValeroEtAl2008}
L.~Gomez-Valero, {\it et~al.\/}, {\it Molecular Ecology\/} {\bf 17}, 4382
  (2008).

\bibitem{ClarkEtAl2007}
A.~G. Clark, {\it et~al.\/}, {\it Nature\/} {\bf 450}, 203 (2007).

\bibitem{Lunter2007b}
G.~Lunter, {\it Bioinformatics\/} {\bf 23}, 289 (2007).

\bibitem{HegerPonting2008}
A.~Heger, C.~Ponting, {\it NAR\/} {\bf 36}, 267 (2008).

\bibitem{ChauxEtAl2007}
N.~de~la Chaux, P.~Messeer, P.~Arndt, {\it BMC Evolutionary Biology\/} {\bf 7}
  (2007).

\bibitem{WangEtAl2009}
Z.~Wang, {\it et~al.\/}, {\it BMC Evol Biol.\/} {\bf 9} (2009).

\bibitem{SacconeEtAl2011}
S.~Saccone, {\it et~al.\/}, {\it Nucleic Acids Res\/}  (2011).

\bibitem{BeissbarthSpeed2004}
T.~Beissbarth, T.~P. Speed, {\it Bioinformatics\/} {\bf 20}, 1464 (2004).

\bibitem{MouseGenome}
{\it Nature\/}  (2002).

\bibitem{MatsenEtAl2010}
F.~A. Matsen, R.~B. Kodner, E.~V. Armbrust, {\it BMC Bioinformatics\/} {\bf
  11}, 538 (2010).

\bibitem{Yang94b}
Z.~Yang, {\it Journal of Molecular Evolution\/} {\bf 39}, 105 (1994).

\bibitem{RannalaYang96}
B.~Rannala, Z.~Yang, {\it Journal of Molecular Evolution\/} {\bf 43}, 304
  (1996).

\bibitem{SiepelHaussler04b}
A.~Siepel, D.~Haussler, {\it Journal of Computational Biology\/} {\bf 11}, 413
  (2004).

\bibitem{YangEtAl2000}
Z.~Yang, R.~Nielsen, N.~Goldman, A.-M. Pedersen, {\it Genetics\/} {\bf 155},
  432 (2000).

\bibitem{ThorneEtAl96}
J.~L. Thorne, N.~Goldman, D.~T. Jones, {\it Molecular Biology and Evolution\/}
  {\bf 13}, 666 (1996).

\bibitem{SiepelHaussler04}
A.~Siepel, D.~Haussler, {\it Molecular Biology and Evolution\/} {\bf 21}, 468
  (2004).

\bibitem{KnudsenHein99}
B.~Knudsen, J.~Hein, {\it Bioinformatics\/} {\bf 15}, 446 (1999).

\bibitem{SiepelHaussler04c}
A.~Siepel, D.~Haussler, {\it Proceedings of the eighth annual international
  conference on research in computational molecular biology, San Diego, March
  27-31 2004\/}, P.~Bourne, D.~Gusfield, eds. (ACM, 2004), pp. 177--186.

\bibitem{PedersenEtAl2006}
J.~S. Pedersen, {\it et~al.\/}, {\it PLoS Computational Biology\/} {\bf 2}, e33
  (2006).

\bibitem{KschischangEtAl98}
F.~R. Kschischang, B.~J. Frey, H.-A. Loeliger, {\it {IEEE} Transactions on
  Information Theory\/} {\bf 47}, 498 (1998).

\bibitem{MohriPereiraRiley2000}
M.~Mohri, F.~Pereira, M.~Riley, {\it Computer Speech and Language\/} {\bf 16},
  69 (2002).

\bibitem{PatenEtAl2008}
B.~Paten, {\it et~al.\/}, {\it Genome Research\/} {\bf 18}, 1829 (2008).

\bibitem{WestessonEtAlArxiv2011}
O.~Westesson, G.~Lunter, B.~Paten, I.~Holmes, {\it arXiv\/}  (2011).
  ArXiv:1103.4347v1.

\bibitem{Moore56}
E.~F. Moore, {\it Gedanken-experiments on Sequential Machines\/} (Princeton
  University Press, Princeton, N.J., 1956), vol.~34 of {\it Annals of
  Mathematical Studies\/}, chap.~5, pp. 129--153.

\bibitem{ThorneEtal91}
J.~L. Thorne, H.~Kishino, J.~Felsenstein, {\it Journal of Molecular
  Evolution\/} {\bf 33}, 114 (1991).

\bibitem{MiklosLunterHolmes2004}
I.~Mikl\'{o}s, G.~Lunter, I.~Holmes, {\it Molecular Biology and Evolution\/}
  {\bf 21}, 529 (2004).

\bibitem{Yang2007}
Z.~Yang, {\it Molecular Biology and Evolution\/} {\bf 24}, 1586 (2007).

\bibitem{MillsEtAl2006}
R.~Mills, {\it et~al.\/}, {\it Genome Research\/} {\bf 16} (2006).

\bibitem{SinhaSiggia2005}
S.~Sinha, E.~Siggia, {\it MBE\/} {\bf 22} (2005).

\bibitem{DoEtAl2004}
C.~B. Do, M.~Brudno, S.~Batzoglou, {PROBCONS}: Probabilistic consistency-based
  multiple alignment of amino acid sequences (2004). Submitted.

\bibitem{DartTutorial}
I.~Holmes, {\it A {DART} tutorial\/}, Berkeley Drosophila Genome Project, LSA
  Room 539, UC Berkeley (2000). A tutorial for probabilistic methods and hidden
  {M}arkov models, presented with the aid of the author's software package
  implementing many common {HMM} algorithms. Available from
  http://www.fruitfly.org/\~{}ihh/.

\end{thebibliography}

\appendix
\section{Simulation parameters and setup}
\applabel{datagen}
\paragraph{Data generation}  Our simulation study is comprised of alignments simulated  using 5 different indel rates (0.005, 0.01, 0.02, 0.04, and 0.08 indels per unit time), each with 3 different substitution rates (0.5, 1, and 2 expected  substitutions per unit time) and 100 replicates.  
Time is defined such that a sequence evolving for time $t$ with substitution rate $r$ is expected to accumulate $rt$ subsitutions per site. 
 We employed an independent third-party simulation program, \indelseqgen, specifically designed to generate realistic protein evolutionary histories \cite{StropeEtAl2009}.  \indelseqgen\ is capable of modeling an empirically-fitted indel length distribution, rate variation among sites, and a neighbor-aware distribution over inserted sequences allowing for small  local duplications.  Since the indel and substitution model used by \indelseqgen\ are separate from (and richer than) those used by \protpal, \protpal\ has no unfair advantage in this test.

\indelseqgen\ v2.0.6 was run with the following command:\\
{\tt cat guidetree.tree| indel-seq-gen -m JTT -u xia --num\_gamma\_cats 3 -a 0.372 --branch\_scale r/b --outfile simulated\_alignment.fa  --quiet --outfile\_format f -s 10000  --write\_anc}

The above command uses the ``JTT'' substition model,  the ``xia'' indel fill model (based on neighbor effects, estimated from E coli k-12 proteins \cite{StropeEtAl2009}), and 3 gamma-distributed rate categories with shape 0.372.  Branch lengths are scaled by the substitution rate for simulation rate $r$, normalized by the inverse of \indelseqgen's underlying substitution rate ($b=1.2$) so as to adhere to the above definition of evolutionary ``time''.   Similarly, indel rates, which are set in the guide tree file {\tt guidetree.tree}, are scaled by $\frac{b}{r}$ so that $t\lambda^*$ insertions/deletions are expected over time $t$ for rate $\lambda^*$.

The root mean squared error (RMSE) for each error distribution was computed as follows:
\begin{equation}
\eqnlabel{RMSE}
RMSE = \sqrt{ \sum_{replicates} (\rateRatio{*}-1)^2 }
\end{equation}

The true tree was made available to all programs which can utilize a tree (\protpal, \prank, \muscle), representing the use case in which the true tree is known (e.g. via the species tree) but the true alignment is unknown.  
We ran simulations on three different phylogenies: a tree of twelve sequenced {\em Drosophila} genomes \cite{ClarkEtAl2007}
and trees from the mammalian and amniotic clades of the OPTIC database.
We here report results for the {\em Drosophila} tree,
which we empirically observe to show trends consistent with, but more pronounced than, those of the mammalian and amniotic trees.
The clearer trends may be due to the {\em Drosophila} tree being larger than the other trees (12 taxa), or having a diverse range of branch lengths (0.001 - 0.59 expected substitutions/site, at the genome-wide average rate).
The simulation data, reconstructions, and analysis scripts  are available from \simulationlink .

\paragraph{Alignment} 
We investigated several  multiple alignment tools \cite{BradleyEtAl2009,DoEtAl2004,Edgar2004b,KatohEtAl2005,LarkinEtAl2007,LoytynojaGoldman2005} in combination with alignment-conditioned reconstruction methods.  
Programs were run with their default settings, with the exception of \prank\ and \muscle.  To specify ancestral inference, the guide tree, and ``insertions opening forever'', \prank\ used the extra options
``{\tt -writeanc -t <treefile> +F}''.
\prank's {\tt -F} option allows insertions to match characters at alignments closer to the root.  This can be a useful heuristic safeguard when an incorrect tree may produce errors in subtree alignments that cannot be corrected at internal nodes closer to the root.  
Since the true guide tree is provided to \prank,  it is safe to treat insertions in a strict phylogenetic manner via the {\tt +F} option.  
For computational efficiency,  \protpal\ was provided with a  \clustalw\ guide alignment.  Any alignment of the sequences can be used as a guide, and we chose \clustalw\ for its general poor performance, so that \protpal\ would gain no unfair advantage by the information contained in the guide alignment. 
\muscle\ was provided the guide tree with the additional option
``{\tt -usetree <treefile>}''.

\begin{description}
\item {\bf Muscle v3.6}\\ {\tt MUSCLE  -in  unaligned.fa  -out  aligned.fa  -usetree  guidetree.tree} 

\item {\bf PRANK v.080820} \\  {\tt PRANK -d=unaligned.fa   -noxml   -realbranches   -writeanc  -o=output\_directory   -t=guidetree.tree   +F}

\item {\bf Clustal v2.03} \\ {\tt clustalw -INFILE=unaligned.fa -OUTFILE=aligned.fa}

\item {\bf ProbCons v1.12 } \\ {\tt probcons unaligned.fa > aligned.fa  } 

\item {\bf FSA v1.08 } \\ {\tt fsa unaligned.fa > aligned.fa  }  

\item {\bf MAFFT v6.818b } \\ {\tt mafft unaligned > aligned.fa } 
\end{description}

\begin{verbatim} 
muscle 3.6 -in <infile> -out <outfile> -usetree <guide tree>
prank v. 
cluswal 2.03 $(clustalw) -INFILE=$< -OUTFILE=$@.clustalw
probcons 1.12 probcons <infile>
fsa 1.08 fsa <infile>
mafft v6.818b mafft <infile>
\end{verbatim} 

\paragraph{Imputing indel histories} The ancestral reconstruction programs \protpal\ and \prank\ were used to directly impute indel histories.  The remaining tools were augmented to reconstruction tools by post-processing their MSAs using the maximum parsimony algorithm described in \cite{SinhaSiggia2005}, with the ambiguous cases described therein (e.g. where a column of characters could be equally parsimoniously explained by a deletion on one child branch or an insertion on the other) resolved by a uniformly random choice from the possible solutions.
Indel rates were estimated by counting indel events in MAP reconstructed histories:
\begin{equation}
\eqnlabel{mapHistoryParams}                      
\hat{\theta}_{\hat{H}} = \argmax_{\theta'} P(\theta'|\hat{H},S,T) = \argmax_{\theta'} P(\hat{H},S|T,\theta')
\end{equation}
where the latter step assumes a flat prior, $P(\theta') = \mbox{const.}$

This statistic is not without its problems.
For one thing, we use an initial guess of $\theta$ to estimate $\hat{H}$.
Furthermore, for an unbiased estimate, we should sum over all histories, rather than conditioning on the MAP reconstructed history.
This summing over histories would, however, require multiple expensive calculations of $P(S|T,\theta)$, where conditioning on $\hat{H}$ requires only one such calculation.  
We further justify our benchmark of parameter estimates conditioned on a MAP-reconstructed history by noting that this the {\em de facto} method employed by large-scale genomics studies focusing on indels \cite{KamnevaEtAl2010, ZhangEtAl2011,ZhuWangEtAl2009, GomezValeroEtAl2008}.

As well as imputing indel rates from reconstructed histories,
we also tried using the \dawglambda\ program in the DAWG package \cite{Cartwright2005}, which estimates indel rates from MSAs directly (without attempting reconstruction).

\paragraph{Estimating substitution rates}  Substitution rates were estimated for each inferred alignment using \xrate's built-in EM algorithm and the following simple rate matrix.
Given an equilibrium distribution over amino acid characters, with $\pi_i$ defining the proportion of character $i$, the rate of character $i$ mutating to $j$ is set to $r \pi_j$ where $r$ is the only free rate parameter.  \xrate's estimate of $r$ is taken to be the average substitution rate of the MSA.

By using \indelseqgen's branch-scale option and changing the indel rate parameters accordingly, we are able to modulate the substitution and indel rates independently in the data generation step.  This true substitution rate and the rate inferred by \xrate\ are then directly comparable.

\section{Supplemental Figures: simulated data analysis  }
\applabel{suppSim}

\begin{center}
\begin{figure}[h!]
\includegraphics[width=1\textwidth]{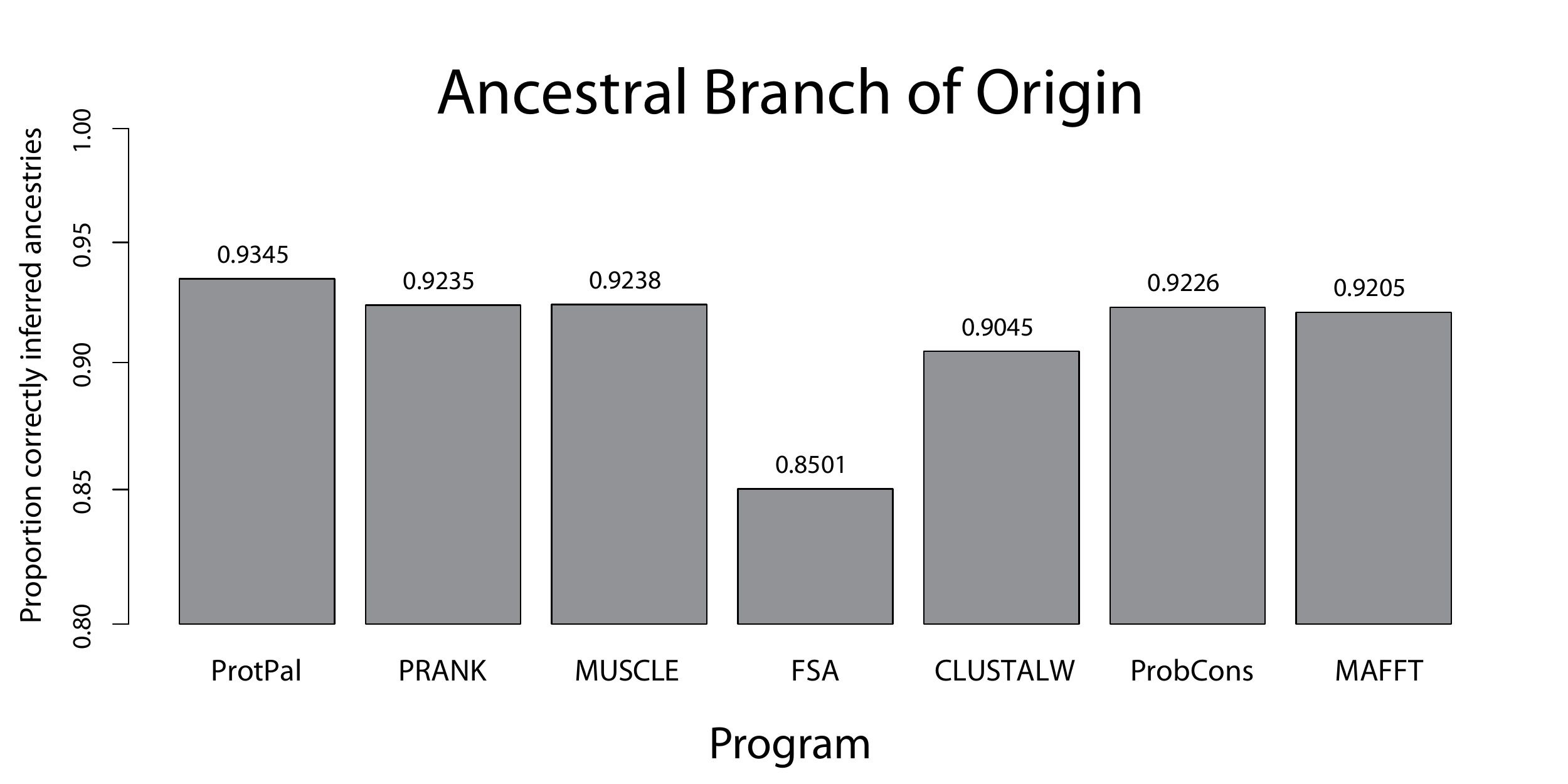}
\caption{
\protpal\ correctly reconstructs the age of more extant residues than any other program tested.  The $y$-axis shows the proportion of extant residues whose point of origin on the phylogenetic tree was correctly pinpointed by the reconstruction. The branch of origin was found by taking the tree node closest to the root containing a non-gap reconstructed character.  All programs except \fsa\ are in the 92\%-94\% range, owing to the fact that many columns (especially at low indel rates) are devoid of indels, making inference of origin trivial (as these columns' origin is pre-root). 
}
\figlabel{ancestry}
\end{figure}
\end{center}

\begin{center}
\begin{figure}[h!]
\includegraphics[width=1\textwidth]{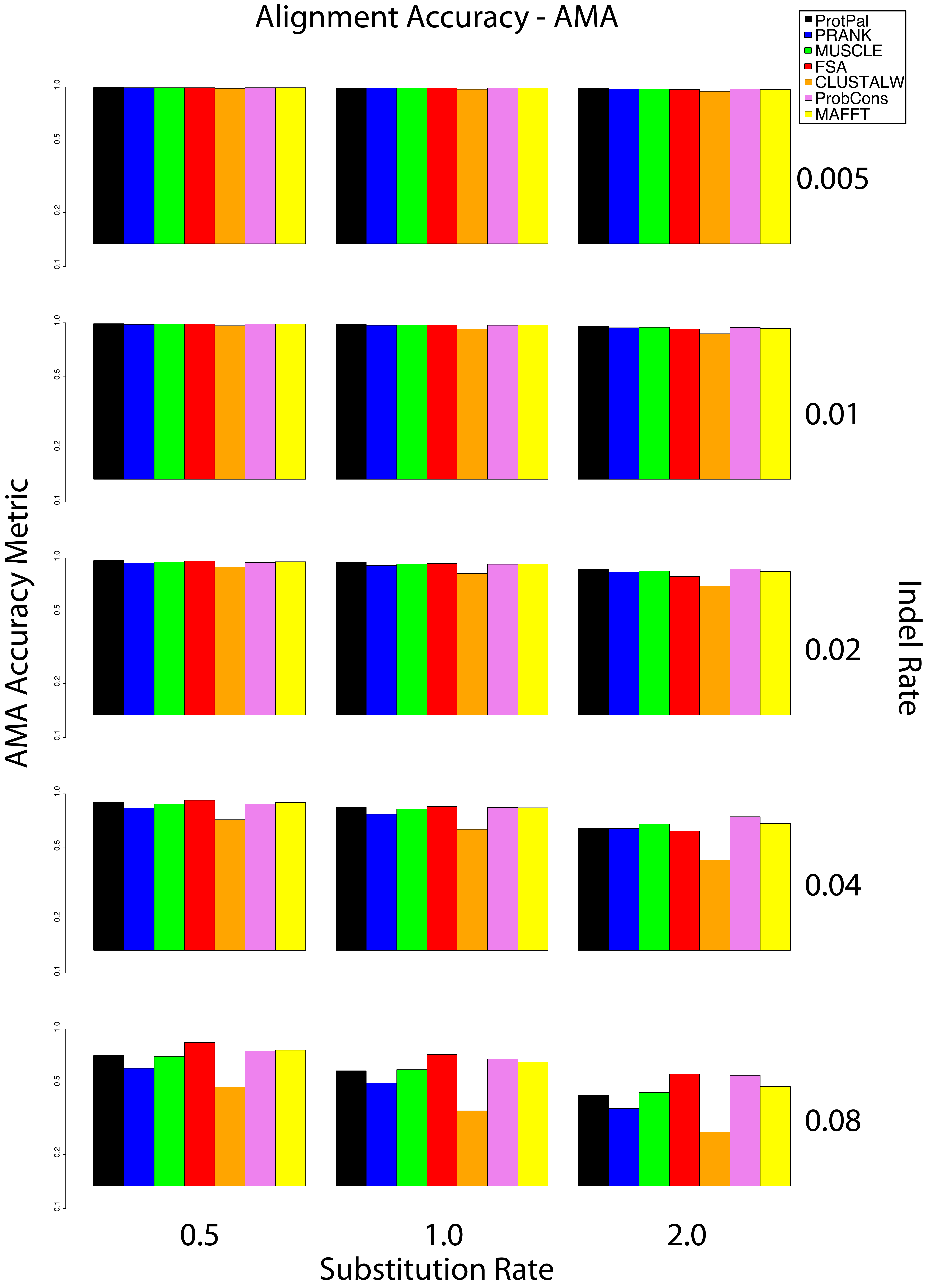}
\caption{
Cross-comparison of AMA scores and rate estimation accuracy reveals that using a single metric to assess alignment accuracy can be unreliable.  AMA scores were computed for each programs alignment of only leaf sequences using {\bf cmpalign} from the DART package \cite{DartTutorial}.
AMA scores are comparable across programs until higher indel rates, where \fsa\ performs best---contrasting with Figures 1 and 2 (main text).  \muscle's accurate deletion rate measurements  at high rates and the low corresponding AMA scores suggest a ``cancellation of biases''.
}
\figlabel{ama}
\end{figure}
\end{center}

\begin{center}
\begin{figure}[h!]
\includegraphics[width=1\textwidth]{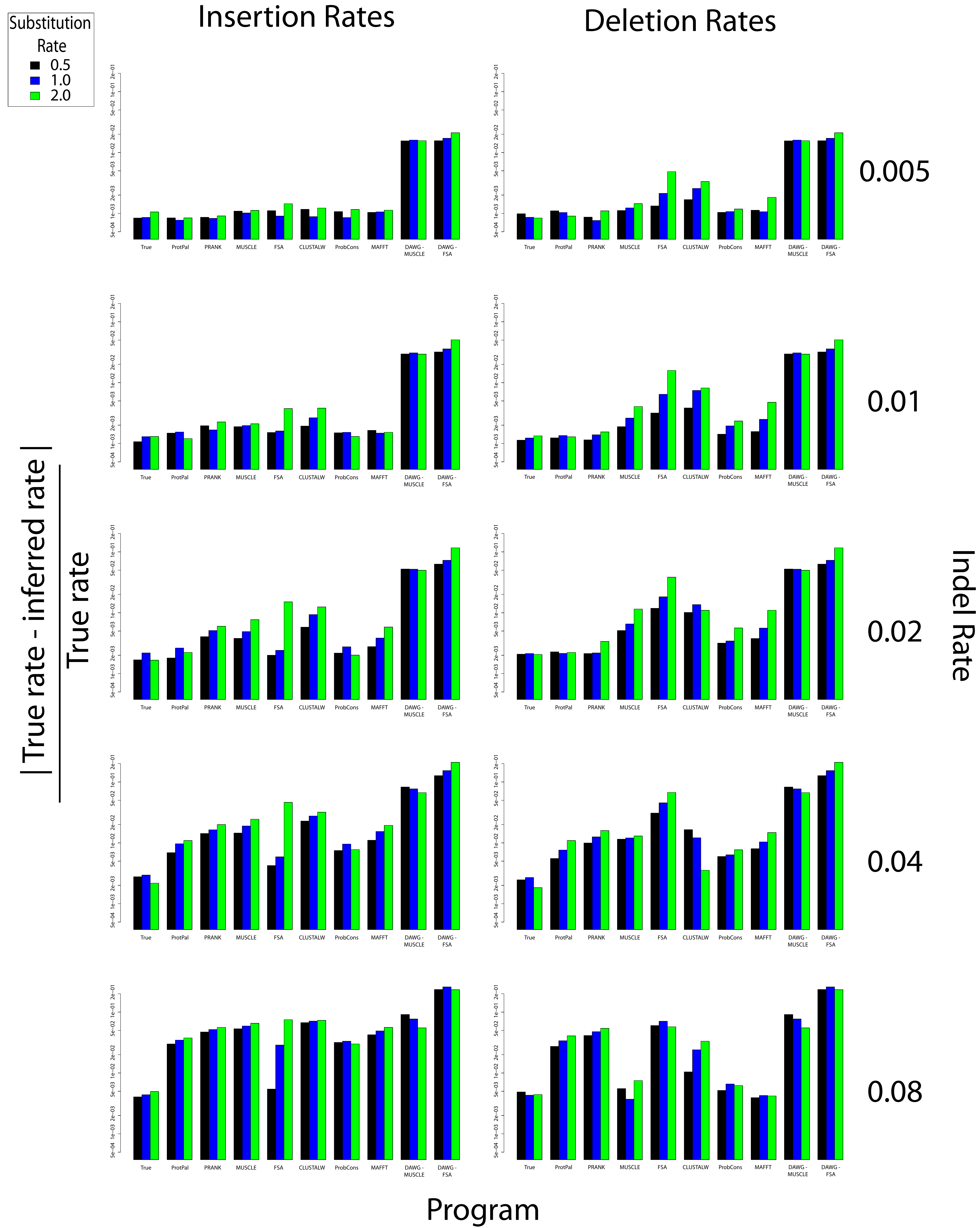}
\caption{
Most programs are relatively robust to variations in the simulated substitution rate, as evidenced by the benchmark data grouped according to substitution rate.  Accuracy of rate estimation is plotted as $|true-inferred|$ on the $y$-axis, with bars grouped by program for each indel rate and 3-tuple of substitution rates.  Higher substitution rates often lead to higher error, presumably because they obscure homologies, making it more difficult to distinguish substitutions from indels.  \fsa\ appears more sensitive to increased substitutions than other programs - at indel rate 0.02, \fsa's insertion rates are as accurate as \protpal's at 0.5 and 1.0 substitutions per site, whereas at the highest substitution  rate (2.0), its error exceeds that of \clustalw. 
}
\figlabel{indelBarPlot_bySub}
\end{figure}
\end{center}

\begin{center}
\begin{figure}[h!]
\includegraphics[width=1\textwidth]{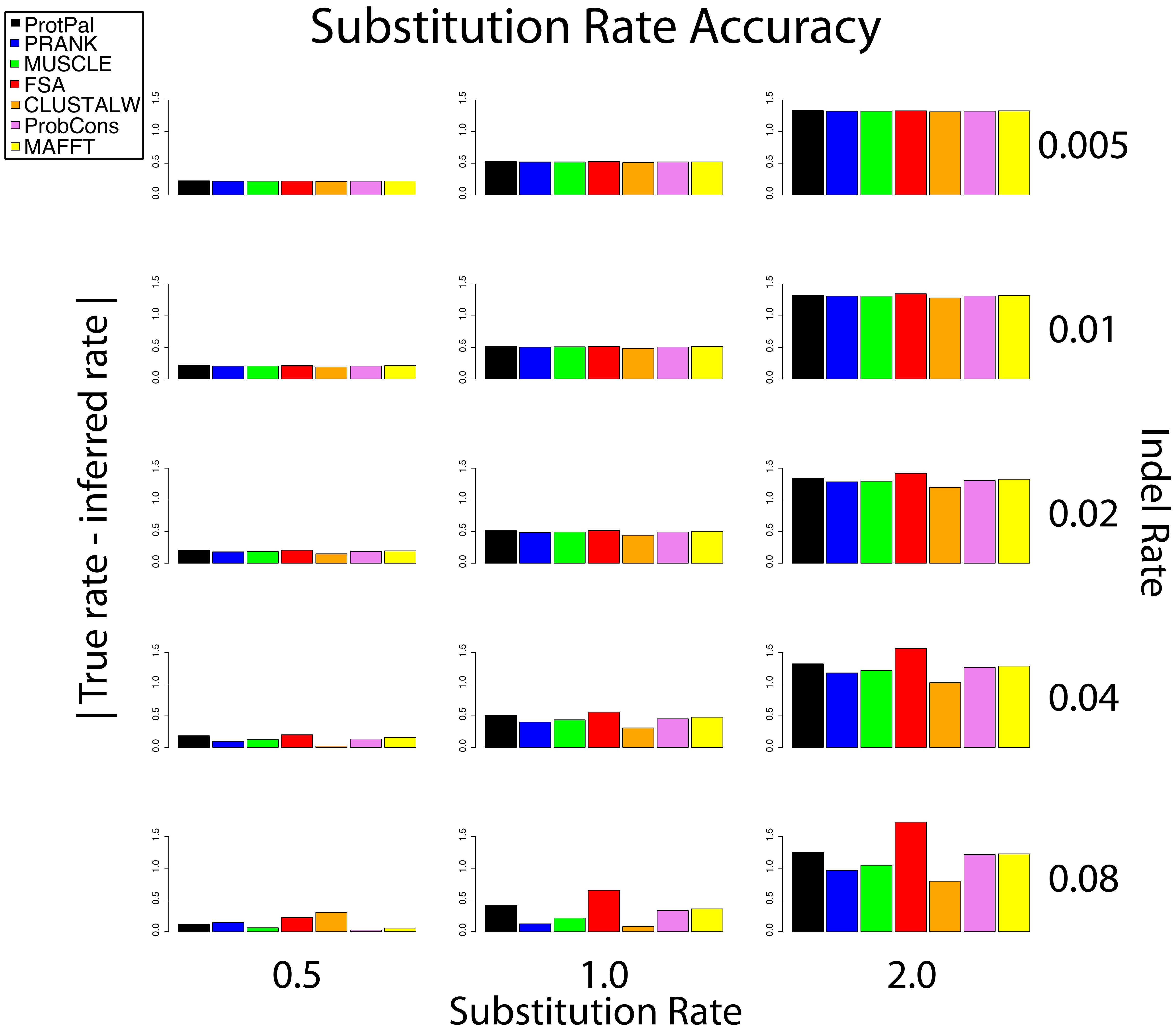}
\caption{
Substitution rates estimated from multiple alignments display comparable accuracy across methods.
}
\figlabel{subRateMatrix}
\end{figure}
\end{center}

\pagebreak
\section{Supplemental Figures: OPTIC analysis  }
\applabel{suppOptic}
In addition to estimating indel rates for all genes in the OPTIC set, we performed various other analyses which were left out of the main text for reasons of space limitations.  We provide figures those displaying results here.

\begin{center}
\begin{figure}[h!]
\includegraphics[width=1\textwidth]{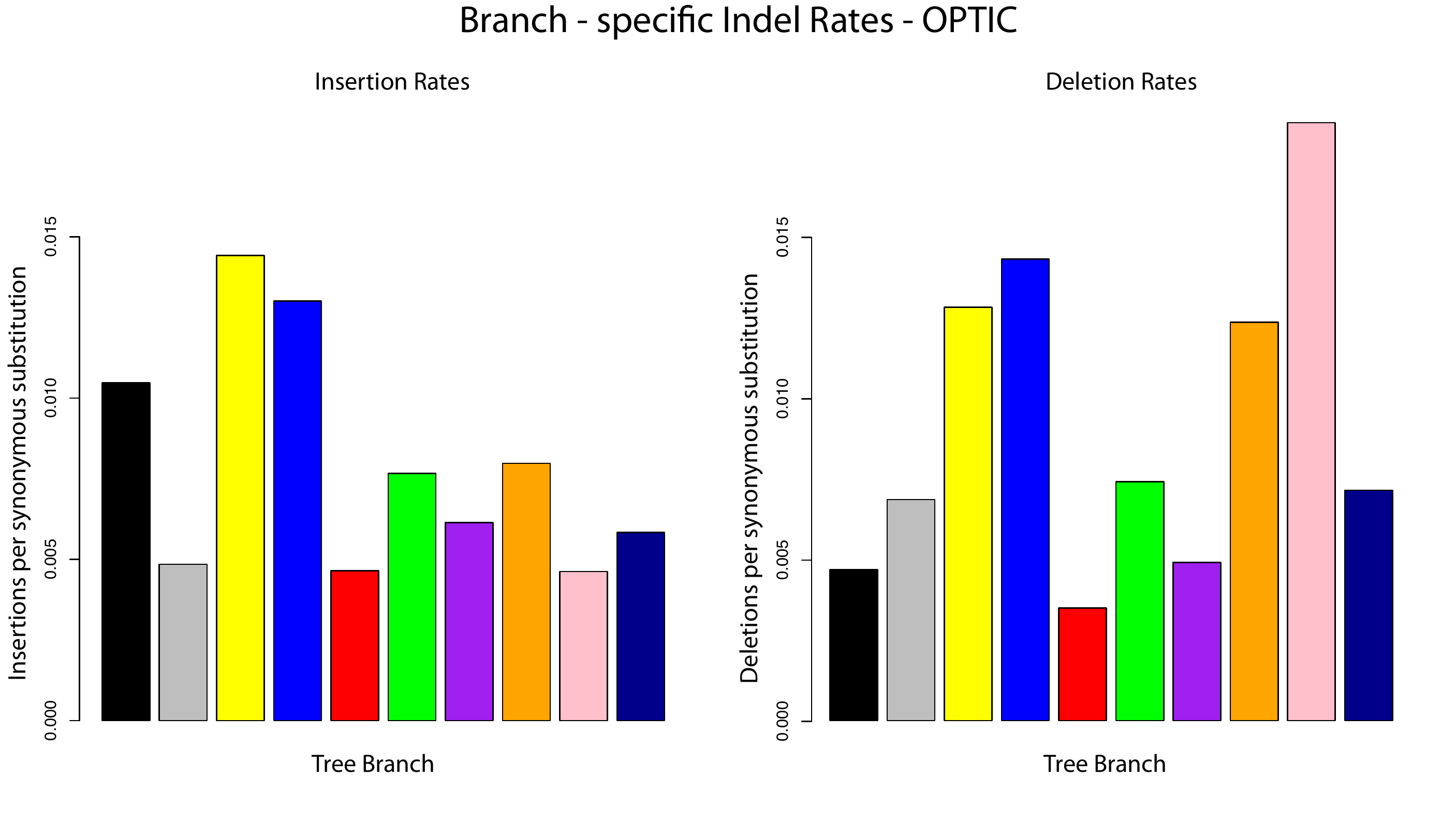}
\caption{Reconstruction allows for estimation of branch-specific indel rates, revealing possibly interesting signals of evolution.  Indel rates were averaged over all alignments, using the species tree shown in \figref{OPTIC.branches}.  The human branch ({\em Euarchontoglires} - H.{\em sapiens}) appears to have experienced unusually many insertions.   The  {\em Amniota - Australophenids} (pink) branch has a higher deletion than insertion rate, though it is difficult to distinguish an insertion on this branch from a deletion on the {\em Amniota} - G.{\em gallus} (navy) branch.  All other branches are comparable between insertions and deletions.  Each bar is  colored according to branches in \figref{OPTIC.branches}. }
\figlabel{OPTIC.branch_indel_rates}
\end{figure}
\end{center}

\begin{center}
\begin{figure}[h!]
\includegraphics[width=1\textwidth]{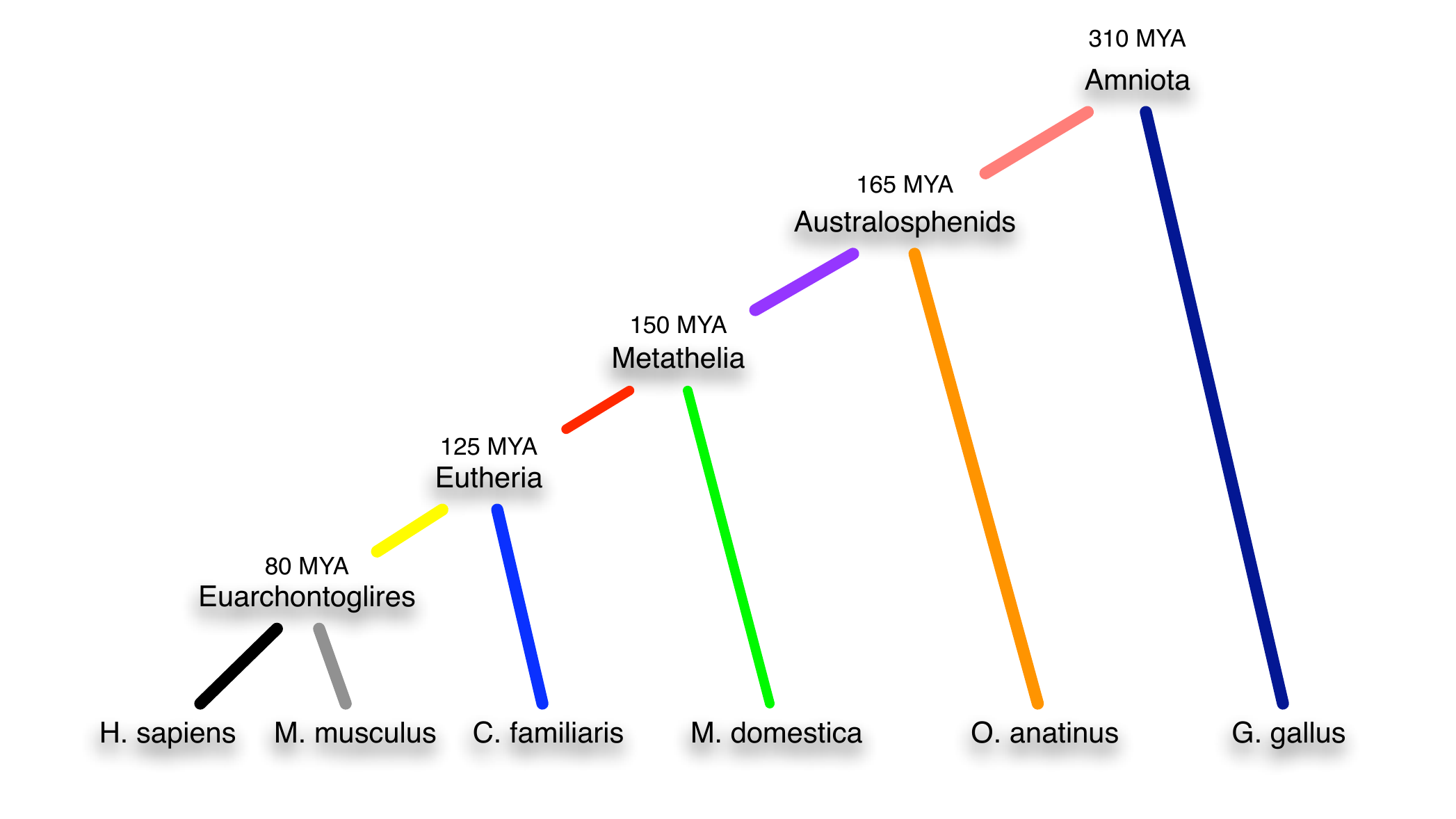}
\caption{The phylogenetic tree used for analysis of OPTIC data, colored to inform the branch-specific Figure \ref{Figures.OPTIC.branch_indel_rates}.}
\figlabel{OPTIC.branches}
\end{figure}
\end{center}

\begin{center}
\begin{figure}[h!]
\includegraphics[width=1\textwidth]{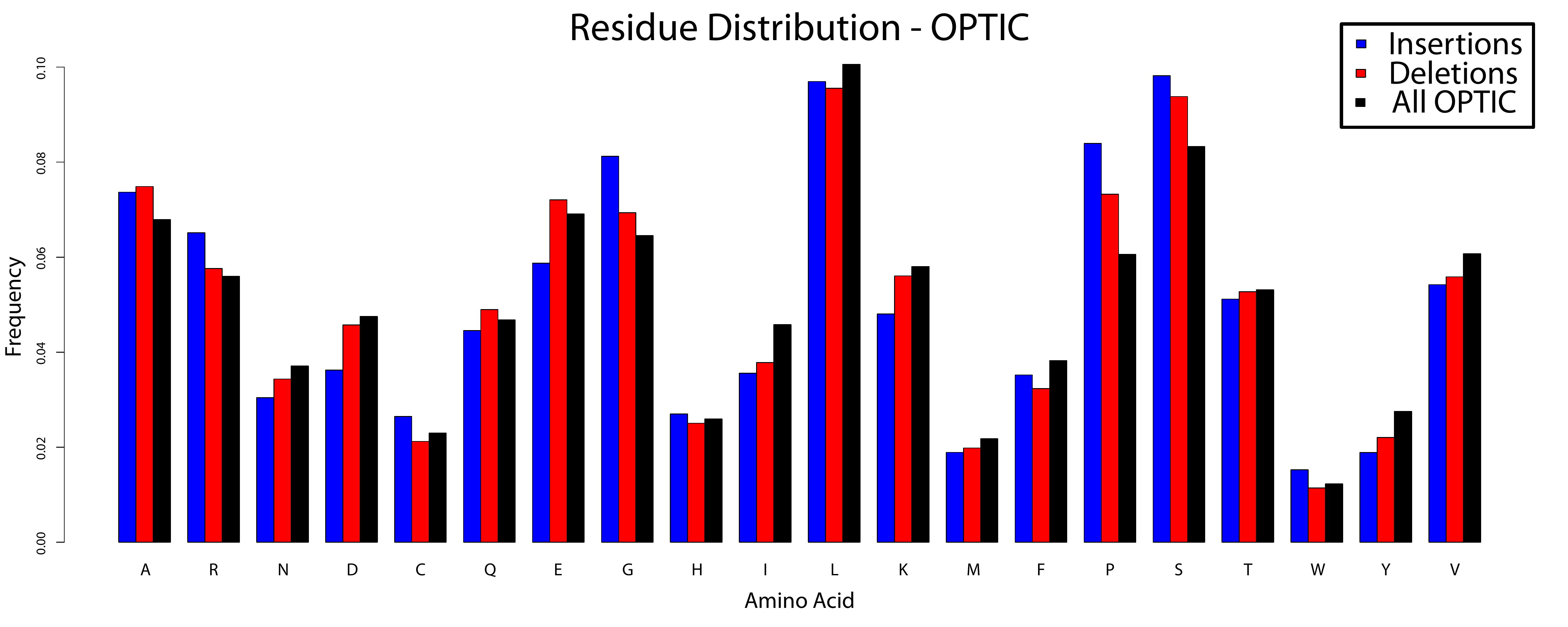}
\caption{Distributions over amino acids are  highly non-uniform, and differ between insertions, deletions, and the overall distribution seen in OPTIC.  Inserted, deleted, and all sequences were separately pooled across all OPTIC genes reconstructed and amino acid distributions were computed for each.}
\figlabel{OPTIC.dist}
\end{figure}
\end{center}

\begin{center}
\begin{figure}[h!]
\includegraphics[width=1\textwidth]{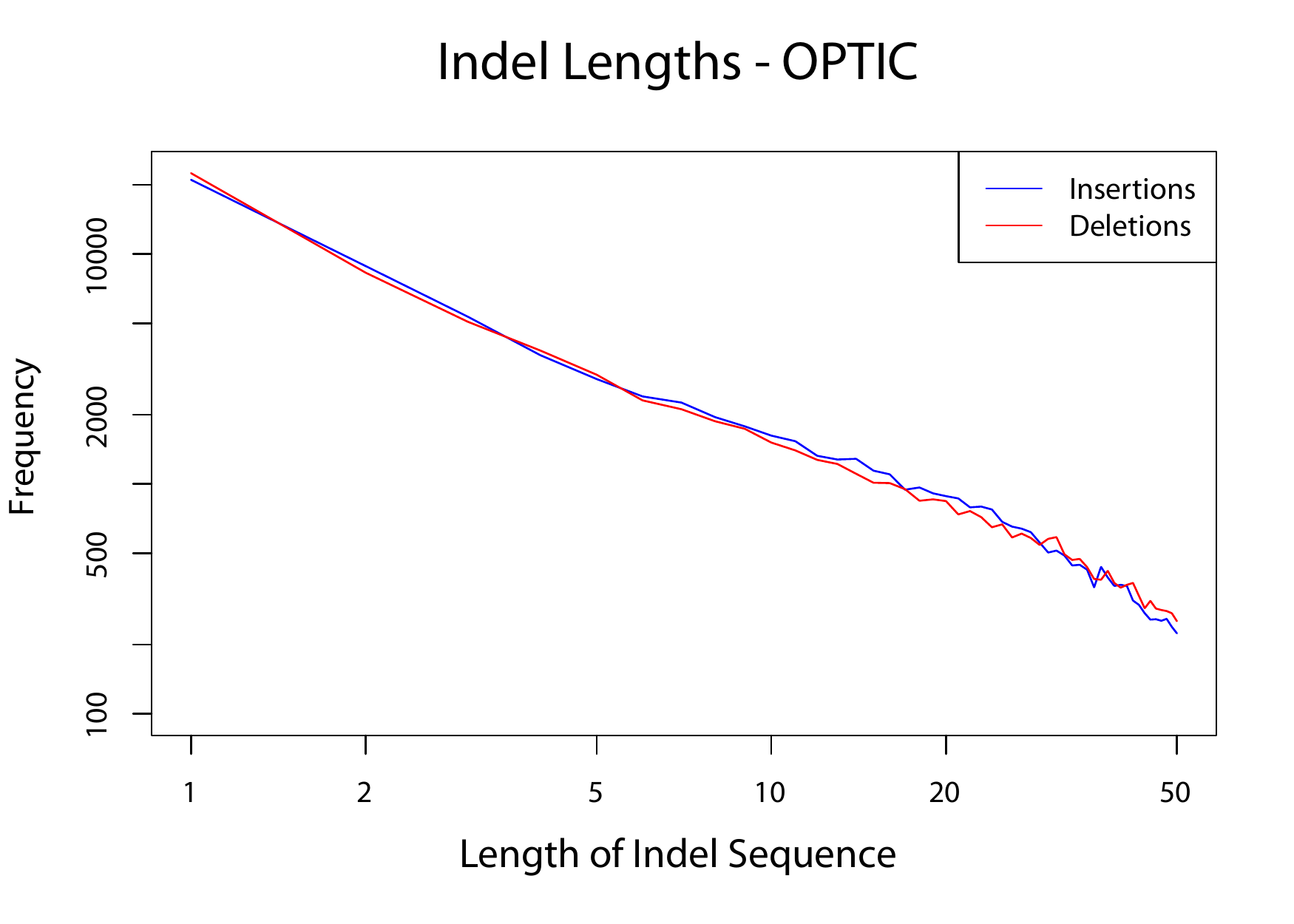}
\caption{
 Lengths of inserted and deleted sequences are similarly distributed, in contrast to the conclusions of previous studies, such as \cite{WangEtAl2009}, which found that  deletions were longer relative to insertions   in C. {\em elegans} sequence data.  While this may represent a genuine difference in the evolution of human and worm genomes, it is likely that the use of deletion-biased aligners (\muscle\ and \clustalw) affected their conclusions.  
}
\figlabel{OPTIC.indel_distribution}
\end{figure}
\end{center}

\begin{center}
\begin{figure}[h!]
\includegraphics[width=1\textwidth]{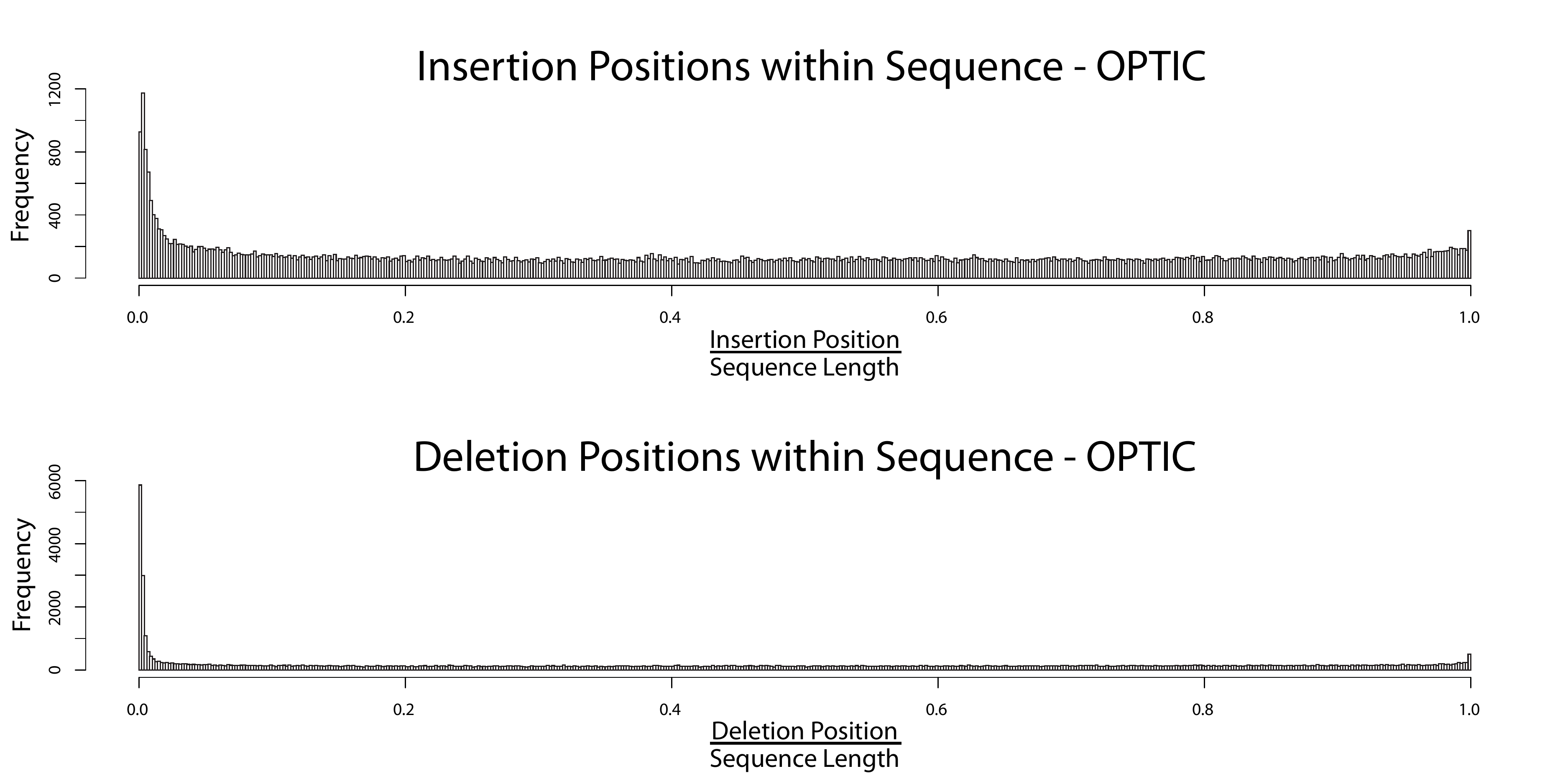}
\caption{
Indels are highly non-uniform in their distribution across genes: we see a 6-fold enrichment for insertions within the N-terminal 1\% of the protein sequence, and a 1.4-fold increase within the C-terminal 1\%.  There is an 14-fold enrichment in deletions within the N-terminal 1\% of the protein sequence, and a 1.8-fold increase within the C-terminal 1\%. 
Indel locations are normalized by gene length to enable combining data across all OPTIC genes analyzed. 
This may be a mix of genuine biology (e.g. indels occur more often near the ends of genes) and artifacts (annotation errors are more likely to occur at the ends of genes). }
\figlabel{OPTIC.indel_positions}
\end{figure}
\end{center}

\begin{center}
\begin{figure}[h!]
\includegraphics[width=1\textwidth]{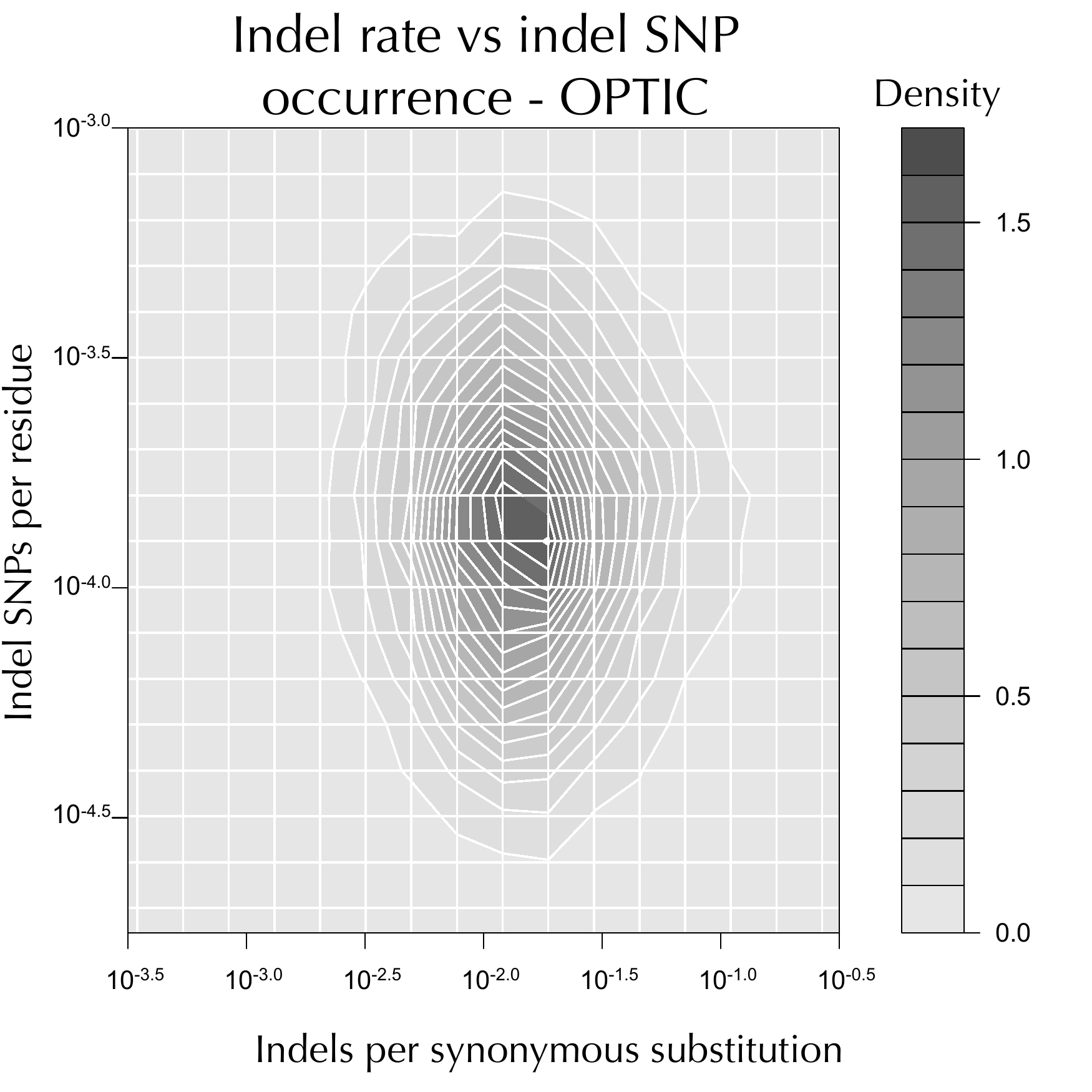}
\caption{Visualizing the number of indel SNPs per residue (using only human sequence) against the evolutionary indel rate (computed across the {\em Amniote} clade) shows no correlation.}
\figlabel{OPTIC.indelVsSNP}
\end{figure}
\end{center}

\end{document}